\documentclass[3p,times,review]{elsarticle}
\usepackage{graphicx}
\usepackage{epstopdf}
\usepackage{epsfig}

\usepackage{tikz}
\usepackage{pgfplots}

\usepackage{fixmath}
\usepackage{xfrac}
\usepackage{amsmath}
\usepackage{amsthm}
\usepackage{stmaryrd}
\usepackage{mathrsfs}

\usepackage{booktabs}
\usepackage{subcaption}

\usepackage[mathlines]{lineno}
\usepackage[breaklinks=true]{hyperref}
\usepackage[noabbrev,nameinlink]{cleveref}
\usepackage{natbib}

\usepackage[ruled,vlined]{algorithm2e}

\usetikzlibrary{arrows.meta}
\pgfplotsset{compat=newest}

\SetSymbolFont{stmry}{bold}{U}{stmry}{m}{n}

\SetKwFor{Loop}{Loop}{}{EndLoop}

\definecolor{JCPLightBlue}{RGB}{5, 86, 165}
\definecolor{JCPOrange}{RGB}{232, 74, 39}

\NewDocumentCommand \dx { O{x} } {\,\mathrm{d} #1}
\NewDocumentCommand \vect { m } { \boldsymbol{#1} }
\NewDocumentCommand \od { m m } { \dfrac{\mathrm{d} #1}{\mathrm{d} #2} }
\NewDocumentCommand \pd { m m } { \dfrac{\partial #1}{\partial #2} }

\NewDocumentCommand \avg { sm } {
    \IfBooleanTF#1
    {\langle #2 \rangle}
    {\left\langle #2 \right\rangle}
}

\NewDocumentCommand \inp { sm } {
    \IfBooleanTF#1
    {\langle #2 \rangle}
    {\left\langle #2 \right\rangle}
}

\NewDocumentCommand \jump { sm } {
    \IfBooleanTF#1
    {\langle #2 \rangle}
    {\left\llbracket #2 \right\rrbracket}
}

\NewDocumentCommand \includefigure { O{1} m } {
    \includegraphics[width=#1\linewidth]{#2}
    \vspace{-15pt}
}

\NewDocumentCommand \newtheoremin { m m m } {
    \newtheorem{#1}{#2}
    \numberwithin{#1}{#3}
}
\newtheoremin{example}{Example}{section}
\newtheoremin{remark}{Remark}{section}
\newtheoremin{definition}{Definition}{section}
\newtheoremin{proposition}{Proposition}{section}
\newtheoremin{lemma}{Lemma}{section}
\newtheoremin{theorem}{Theorem}{section}

\crefname{algocf}{algorithm}{algorithms}
\Crefname{algocf}{Algorithm}{Algorithms}

\journal{Journal of Computational Physics}
\begin{document}

\begin{frontmatter}

\title{Adjoint-based Control of Three Dimensional Stokes Droplets}

\author[uiucaddress]{Alexandru Fikl\corref{correspondingauthor}}
\cortext[correspondingauthor]{Corresponding Author}
\ead{fikl2@illinois.edu}

\author[uiucaddress]{Daniel J. Bodony}

\address[uiucaddress]{
    Department of Aerospace Engineering,
    University of Illinois Urbana-Champaign,
    Urbana, IL 61801, United States}

\begin{abstract}
We develop a continuous adjoint formulation and implementation for controlling
the deformation of clean, neutrally buoyant droplets in Stokes flow through
farfield velocity boundary conditions. The focus is on dynamics where surface
tension plays an important role through the Young-Laplace law. To perform the
optimization, we require access to first-order gradient information, which we
obtain from the linearized sensitivity equations and their corresponding adjoint
by applying shape calculus to the space-time tube formed by the interface
evolution. We show that the adjoint evolution equation can be efficiently
expressed through a scalar adjoint transverse field. The optimal control problem
is discretized by high-order boundary integral methods using Quadrature by Expansion
coupled with a spherical harmonic representation of the droplet surface geometry.
We show the accuracy and stability of the scheme on several tracking-type control
problems.
\end{abstract}


\end{frontmatter}

\section{Introduction}
\label{sc:introduction}

In many droplet-based microfluidic processes and applications, the precise
shape and position of the droplets over time plays a significant role in the
performance and efficiency of the system. A standard simplified model for such
problems consists of the two-phase Stokes equations coupled with interfacial
forces (surface tension) or additional surfactants and gravity (several concrete
examples can be found in~\cite{Pozrikidis2001}). Optimizing aspects of this
type of system is our focus. However, other types of interface evolution share
similar features, e.g. mean curvature flows~\cite{Huisken1984, Deckelnick2005},
fluid-structure interactions or free surface flows, and can be analyzed by
similar methods.

In general, the application of the continuous or discrete adjoint method to
the field of optimal control has been very successful, with applications to
fluid mechanics starting from~\cite{Jameson1988}. However, applications to
two-phase flows are rare due, in part, to the fact that the field itself
remains an active area of research. Phase-field-type models present a compelling
starting point, as the fields representing the flow quantities are smooth
and classic approaches to adjoint methods can be applied. For example,
in~\cite{Hintermuller2014, Garcke2019, Deng2017} the Cahn--Hilliard equations are
coupled with the incompressible Navier--Stokes equations and the corresponding
optimal control problem is solved. On the other-hand, sharp interface
models (see~\cite{Prosperetti2009}) are difficult to handle due to the
discontinuities of the state variables and material quantities. Furthermore,
incorporating interface jump conditions containing higher-order derivatives
of the geometry (mainly curvature) is problematic, see~\cite{Popinet2009}.
Work on two-phase problems has also advanced in recent years, with applications to
fluid-structure interactions~\cite{Feppon2019}, two-phase Stefan
equations~\cite{Bernauer2011}, free-surface flows~\cite{Repke2011, Palacios2012},
geometric flows~\cite{Laurain2015, Laurain2021}, etc. Initial extensions to the
incompressible two-phase Navier--Stokes with Volume-of-Fluid formulation are
presented in~\cite{Diehl2020, Kuhl2021}.

In this paper, we extend the work presented initially in~\cite{Fikl2021}.
The focus of the previous work was on the optimal control of a single
droplet under axisymmetric assumptions. We relax these assumptions and
extend the applications to fully three-dimensional problems with multiple
droplets. While applications to scenarios of practical interest,
where hundreds or thousands of droplets are present in the system, are still
prohibitively expensive, we present several improvements. First, we show that
the vector adjoint evolution equation from~\cite{Fikl2021} can be transformed
into a scalar evolution equation for the normal component only. This is a
significant reduction in cost, especially in three dimensions. Furthermore,
we make use of state-of-the art methods for solving boundary integral equations
to improve the general performance of the Stokes solver itself. In this work,
the Quadrature by Expansion method~\cite{Klockner2013} is used, for which
accurate Fast Multipole Methods have been developed~\cite{Wala2019}. This
allows constructing a fast solver that scales approximately linearly in the
number of degrees of freedom.

The outline of the paper is as follows. We start in~\Cref{sc:control} with
a description of the two-phase control problem, focusing on the constraints
and the cost functional. In~\Cref{sc:gradient}, we present the main ideas
behind applying shape calculus to PDEs on moving domains. These
concepts are applied to deriving the linearized and adjoint equations for
the control problem. In~\Cref{sc:methods}, we present a discretization of the
state and adjoint systems through a coupled representation by boundary integral
methods (for the Stokes problem) and spherical harmonics (for the geometry).
Finally, \Cref{sc:results} presents several verification tests and applications
to three-dimensional multi-droplet systems. We conclude with some remarks
in~\Cref{sc:conclusions}.

\section{Control of Two-Phase Stokes Flow}
\label{sc:control}

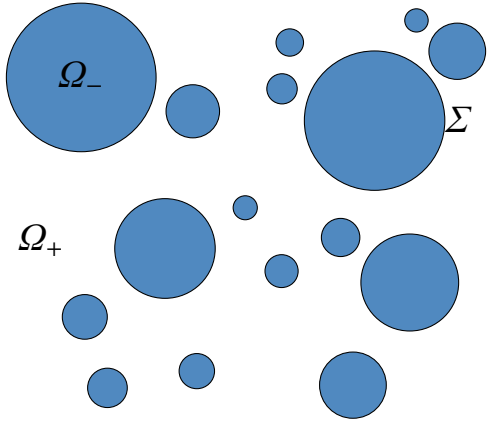
\begin{figure}[ht!]
\centering
\begin{tikzpicture}
\draw[fill=JCPLightBlue!70] (4.351, 1.434) circle [radius=0.643];
\draw[fill=JCPLightBlue!70] (3.888, 3.580) circle [radius=0.924];
\draw[fill=JCPLightBlue!70] (2.186, 2.425) circle [radius=0.159];
\draw[fill=JCPLightBlue!70] (0.028, 4.153) circle [radius=0.985];
\draw[fill=JCPLightBlue!70] (1.496, 3.704) circle [radius=0.352];
\draw[fill=JCPLightBlue!70] (2.772, 4.616) circle [radius=0.181];
\draw[fill=JCPLightBlue!70] (1.130, 1.885) circle [radius=0.661];
\draw[fill=JCPLightBlue!70] (0.373, 0.037) circle [radius=0.261];
\draw[fill=JCPLightBlue!70] (3.603, 0.073) circle [radius=0.439];
\draw[fill=JCPLightBlue!70] (0.075, 0.978) circle [radius=0.296];
\draw[fill=JCPLightBlue!70] (3.441, 2.031) circle [radius=0.253];
\draw[fill=JCPLightBlue!70] (1.549, 0.259) circle [radius=0.233];
\draw[fill=JCPLightBlue!70] (2.664, 1.585) circle [radius=0.218];
\draw[fill=JCPLightBlue!70] (4.976, 4.500) circle [radius=0.373];
\draw[fill=JCPLightBlue!70] (4.439, 4.909) circle [radius=0.154];
\draw[fill=JCPLightBlue!70] (2.670, 4.002) circle [radius=0.200];

\node at (0.028130253396964622, 4.153107184856285) {\Large $\Omega_-$};
\node at (-0.5, 2) {\Large $\Omega_+$};
\node at (5, 3.6) {\Large $\Sigma$};
\end{tikzpicture}

\caption{Example droplet configuration.}
\label{fig:droplets}
\end{figure}

In this paper we will consider the evolution of a system of droplets in an
otherwise infinite domain. To this end, we define an open finite domain
$\Omega_- \subset \mathbb{R}^d$, where $d = 3$, and its complement
$\Omega_+ \triangleq \mathbb{R}^d \setminus \bar{\Omega}_-$ to denote the
interior of the droplets and the surrounding fluid, respectively. The droplet
surface $\Sigma \triangleq \bar{\Omega}_+ \cap \bar{\Omega}_+$ is assumed to
be a finite set of disjoint, closed, bounded, and orientable surfaces at each
instance of time (see~\Cref{fig:droplets}).

The interface $\Sigma(t)$ is the focus in the control of two-phase flows.
As such, we define $\mathcal{D} \subset \mathbb{R}^d$ as a bounded open set
which will contain all admissible interface configurations $\Sigma(t)$
for our problem. The general optimization problem we will be looking at
pertains to evolutions equations of the form
\begin{equation} \label{eq:evolution}
\begin{cases}
\dot{\vect{X}} = \vect{V}(t, \vect{X}(t), \vect{g}),
    & \quad t \in [0, T], \\
\vect{X}(0) = \vect{X}_0,
\end{cases}
\end{equation}
where $\vect{X}(t) \in \mathcal{D}$ is a parametrization of the interface
$\Sigma(t)$ and $\vect{g}$ represents a chosen control. For sufficiently smooth
right-hand sides, we can define a transformation $\vect{T}(\vect{V}): \mathcal{D}
\to \mathcal{D}$ by $\vect{X}(t) = \vect{T}(\vect{V})(\vect{X}_0)$ for every
flow $\vect{V}$. We can then follow the results from~\cite{Moubachir2006} to
derive evolution equations for the linearized problem, the adjoint problem and
ultimately an expression for the adjoint-based gradient. This will be discussed
in detail in~\Cref{sc:gradient}.

\subsection{Quasi-static Two-Phase Stokes Flow}
\label{ssc:control:stokes}

We consider the quasi-static evolution of multiple droplets in a viscous
incompressible flow. The droplets experience no phase change, but are driven
by surface tension forces at the fluid-fluid interface. In each phase, the
fluid is described by the velocity fields $\vect{u}_\pm: \Omega_\pm \to \mathbb{R}^d$
and the pressures $p_\pm: \Omega_\pm \to \mathbb{R}_+$. Assuming a low-Reynolds
number, the equations governing the flow are the two-phase Stokes
equations
\begin{equation} \label{eq:stokes}
\begin{cases}
\nabla \cdot \vect{u}_\pm = 0,
    & \quad \vect{x} \in \Omega_\pm(t), \\
\nabla \cdot \sigma_\pm[\vect{u}, p] = 0,
    & \quad \vect{x} \in \Omega_\pm(t), \\
\vect{u}_+ \to \vect{u}_\infty(\vect{g}),
    & \quad \|\vect{x}\| \to \infty,
\end{cases}
\end{equation}
where $\vect{u}_\infty$ defines an intended decay at infinity of the
velocity field $\vect{u}_+$. The full velocity field $\vect{u}$ is to be
understood as a superposition of $\vect{u}_\infty$ and a flow that decays to zero
in the farfield. The Cauchy stress tensor $\sigma_\pm$ and the rate of strain
tensor $\varepsilon_\pm$ are written as
\[
\begin{aligned} \label{eq:stress}
\sigma_\pm[\vect{u}, p] \triangleq \,\, &
    -p_\pm I + 2 \mu_\pm \varepsilon_\pm[\vect{u}], \\
\varepsilon_\pm[\vect{u}] \triangleq \,\, &
\frac{1}{2} (\nabla \vect{u}_\pm + \nabla \vect{u}_\pm^T),
\end{aligned}
\]
where $\mu_\pm \in \mathbb{R}_+$ are the dynamic viscosities of each fluid.
We will assume that the surface forces are entirely due to a constant surface
tension. The boundary conditions at the drop surface are the continuity
of velocity and the Young-Laplace law, i.e.
\begin{equation} \label{eq:jumps}
\begin{cases}
\jump{\vect{u}} = 0, \\
\jump{\vect{n} \cdot \sigma[\vect{u}, p]} = \gamma \kappa \vect{n},
\end{cases}
\end{equation}
where $\gamma$ is a constant surface tension coefficient, $\kappa$ is the
total curvature (sum of the principal curvatures) and $\vect{n}$ is the
exterior normal to $\Omega_-$. The jump at the interface is
defined as
\[
\jump{f} \triangleq f(\vect{x}_+) - f(\vect{x}_-),
\quad \text{where }
f(\vect{x}_\pm) \triangleq \lim_{\epsilon \to 0^+} f(\vect{x} \pm \epsilon \vect{n}).
\]

The equations are non-dimensionalized by making use of a characteristic
velocity magnitude $U$ and a characteristic droplet radius $R$. The resulting
non-dimensional parameters are the viscosity ratio $\lambda$ and the
Capillary number $\mathrm{Ca}$,
\[
\lambda \triangleq \frac{\mu_-}{\mu_+}
\quad \text{and} \quad
\mathrm{Ca} \triangleq \frac{\mu_+ U}{\gamma},
\]

The non-dimensional equations will be used going forward. In this form,
non-dimensional weighted jump and average operators are defined as
\[
\jump{f}_\lambda \triangleq f(\vect{x}_+) - \lambda f(\vect{x}_-)
\quad \text{and} \quad
\avg{f}_\lambda \triangleq \frac{1}{2} (f(\vect{x}_+) + \lambda f(\vect{x}_-)).
\]

In order to evolve the interface in time, we use a quasi-static approach
in which we first compute the velocity from~\eqref{eq:stokes} and then
displace the interface using the ODE from~\eqref{eq:evolution}.
For Stokes flow, the motion law is given by
\begin{equation} \label{eq:kinematic}
\vect{V}(\vect{u}, \vect{X}) \triangleq
    (\vect{u} \cdot \vect{n}) \vect{n}
    + (I - \vect{n} \otimes \vect{n}) \vect{w},
\end{equation}
which is well-defined since $\jump{\vect{u}} = 0$ allows for a unique
interface velocity field. At a continuous level, only the normal component
of the velocity is necessary to define the deformation of the interface.
Tangential components from $\vect{w}: [0, T] \times \mathbb{R}^3 \to \mathbb{R}^3$
can be seen as reparametrizations and represent the same abstract shape.
They  become important in the discretization stage, as discussed
in~\Cref{sc:methods}, and should be included in the adjoint to obtain accurate
gradients.

\subsection{Cost Functionals}
\label{ssc:control:costs}

Our goal is to find controls $\vect{g}$ (from~\eqref{eq:evolution}) that minimize
a class of tracking-type cost functionals. The general form of the cost
functionals we will be looking at is
\begin{equation} \label{eq:cost}
J(\vect{g}) \triangleq
\frac{\alpha_1}{2} \int_0^T \int_{\Sigma(t)}
    (\vect{u} \cdot \vect{n} - u_d)^2 \dx[S] \dx[t]
+ \frac{\alpha_2}{2} \int_0^T \|\vect{x}_c(t) - \vect{x}_d(t)\|^2 \dx[t]
+ \frac{\alpha_3}{2} \|\vect{x}_c(T) - \vect{x}_{d, T}\|^2,
\end{equation}
where $u_d$ is a target surface normal velocity field, while $\vect{x}_d(t)$
and $\vect{x}_{d, T}$ are target droplets centroids. We mainly consider the
droplet centroids as a way to describe the position, but any quantity
that is invariant to reparametrizations can be used. This assumption can be
relaxed by considering reparametrizations as described in~\cite{Luft2020}. Then,
we consider the optimization problem
\begin{equation} \label{eq:optim}
\begin{cases}
\min J(\vect{g}), \\
\text{subject to~\eqref{eq:evolution} (with~\eqref{eq:kinematic}),
\eqref{eq:stokes} and~\eqref{eq:jumps}.}
\end{cases}
\end{equation}

The main control variable we will be focusing on are the farfield boundary
conditions $\vect{u}_\infty(\vect{g})$. For finite domains, this is a
straightforward choice. However, as the farfield boundary conditions are
used to impose a decay in the velocity field, further discussion is required
(see~\Cref{sc:gradient}).

In the absence of an evolution equation, such as~\eqref{eq:evolution},
we can consider a static system. In this case, the interface $\vect{X}$
becomes the control and we are left with a standard shape optimization
problem. As shown in~\cite{Fikl2021}, the static problem is an important
stepping stone in deriving and verifying the required optimality conditions.
We will be making use of the static problem to verify the adjoint equations
for the three-dimensional problem presented here as well.

\section{Cost Gradients}
\label{sc:gradient}

\begin{figure}[ht!]
\centering
\begin{tikzpicture}[scale=0.65]
\begin{scope}[opacity=0.5]
\draw[fill=black!10] (-4, 0) arc (180:360:4cm and 2cm);
\draw[dashed, fill=black!10] (-4, 0) arc (180:0:4cm and 2cm);
\draw[fill=black!10] (-4, 5) arc (180:360:4cm and 2cm);
\draw[dashed, fill=black!10] (-4, 5) arc (180:0:4cm and 2cm);
\draw[fill=black!10] (0, 10) ellipse [x radius=4, y radius=2];
\draw (-4, 0) -- (-4, 10);
\draw (+4, 0) -- (+4, 10);
\end{scope}

\draw[thick, red, fill=JCPLightBlue!40, opacity=50] (-1.5, 0) arc (180:360:1.5cm and 0.5cm);
\draw[thick, dashed, red, fill=JCPLightBlue!40, opacity=50] (-1.5, 0) arc (180:0:1.5cm and 0.5cm);
\draw[thick, red, fill=JCPLightBlue!40, opacity=50] (-1, 5) arc (180:360:1cm and 0.5cm);
\draw[thick, dashed, red, fill=JCPLightBlue!40, opacity=50] (-1, 5) arc (180:0:1cm and 0.5cm);
\draw[thick, red, fill=JCPLightBlue!40] (0, 10) ellipse [x radius=1.75, y radius=0.5];

\draw[thick, red] (-1.5, 0) .. controls (-0.8, 5) .. (-1.75, 10);
\draw[thick, red] (1.5, 0) .. controls (0.8, 5) .. (1.75, 10);

\begin{scope}[xshift=-6.5cm]
\draw[ultra thick, -latex] (0, 0) -- (1, 0) node [below] {$x$};
\draw[ultra thick, -latex] (0, 0) -- (0, 1) node [left] {$t$};
\draw[ultra thick, -latex] (0, 0) -- (-0.7, -0.7) node [above] {$y$};
\end{scope}

\node at (0, 0) {$\Omega_-(0)$};
\node at (0, 5) {$\Omega_-(t)$};
\node at (0, 10) {$\Omega_-(T)$};
\node at (0, -1) {$\Sigma(0)$};
\node at (0, 4) {$\Sigma(t)$};
\node at (0, 9) {$\Sigma(T)$};
\node at (0, 2.5) {$\Omega_-(\mathbf{V})$};
\node at (1.75, 2.5) {$\Sigma(\mathbf{V})$};
\end{tikzpicture}

\caption{
    Schematic of the (two-dimensional) $\Omega(\vect{V})$ and $\Sigma(\vect{V})$
    space-times tubes and transverse slices $\Omega(t)$ (blue) and $\Sigma(t)$
    (red) at fixed times $t$. The perturbations $\tilde{\vect{V}}$ only act
    in the transverse direction (gray slices).}
\label{fig:tubes}
\end{figure}
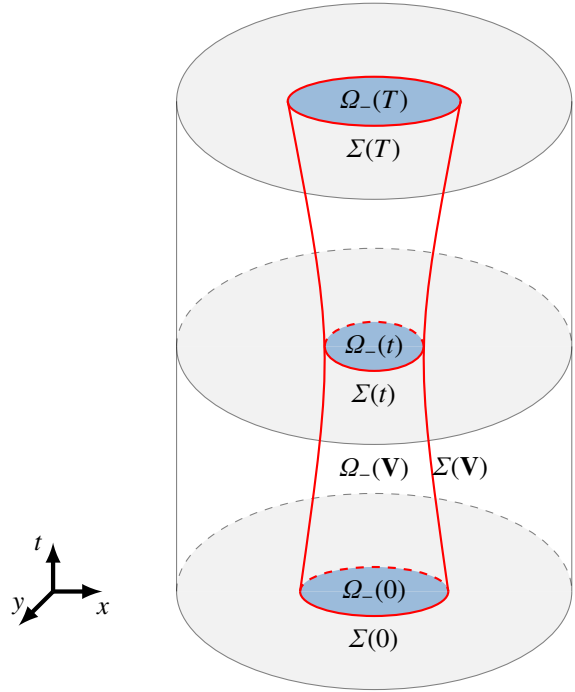

To understand the optimization problem proposed by~\eqref{eq:optim}, we must
introduce several important notions from shape calculus and more general
perturbations of moving domains. These ideas are detailed in~\cite{Moubachir2006}.
First, for an initial configuration $\Omega_0$ and interface $\Sigma_0$, we
define the space-time tubes on $\mathbb{R}_+ \times \mathbb{R}^d$, corresponding
to a velocity field $\vect{V}$, by (see also~\Cref{fig:tubes})
\[
\begin{aligned}
\Omega(\vect{V}) \triangleq \,\, &
    \bigcup_{t \in [0, T]} \{t\} \times \Omega(t)
    = \bigcup_{t \in [0, T]} \{t\} \times \vect{T}(\vect{V})(\Omega_0), \\
\Sigma(\vect{V}) \triangleq \,\, &
    \bigcup_{t \in [0, T]} \{t\} \times \Sigma(t)
    = \bigcup_{t \in [0, T]} \{t\} \times \vect{T}(\vect{V})(\Sigma_0).
\end{aligned}
\]

A perturbation in the control $\vect{g}$ can then be seen as giving rise
to a perturbed space-time tube $\Omega_\epsilon(\vect{V})$. Perturbations
obtained in this way are called \emph{transverse} because they occur only in
``horizontal'' slices of each $\Omega(t)$ domain for fixed $t$. This allows
recovering most of the classic formulae from shape calculus in the context
of moving domains. To clarify these ideas, we consider a perturbation of the
control and the corresponding perturbation of the motion law, i.e.
\[
\begin{aligned}
\vect{g}_\epsilon = \,\, & \vect{g} + \epsilon \tilde{\vect{g}}, \\
\vect{V}_\epsilon(t, \vect{x}; \vect{g}) = \,\, &
    \vect{V}(t, \vect{x}; \vect{g})
    + \epsilon \vect{W}(t, \vect{x}; \vect{g}, \tilde{\vect{g}}),
\end{aligned}
\]
for a specific $\vect{W}$ such that $\vect{W}(0, \vect{x}; \vect{g},
\tilde{\vect{g}}) = \vect{0}$. The expression for $\vect{V}_\epsilon$ is valid
under sufficient regularity assumptions and can be used to obtain
first-order perturbations of the tube $\Omega(\vect{V})$. To obtain derivatives
of the cost functional~\eqref{eq:cost}, we are interested in functionals of the
form
\[
\begin{aligned}
J_1(\vect{V}) \triangleq \,\, & \int_0^T \int_{\Omega(t)}
    f(\vect{V}) \dx[V] \dx[t], \\
J_2(\vect{V}) \triangleq \,\, & \int_0^T \int_{\Sigma(t)}
    g(\vect{V}) \dx[S] \dx[t],
\end{aligned}
\]
where $f, g: \Omega(t) \to \mathbb{R}$. From~\cite[Theorem 5.4, 5.5]{Moubachir2006},
we have that their directional derivatives are given by
\begin{equation} \label{eq:shape_derivatives}
\begin{aligned}
D J_1(\vect{V})[\vect{W}] = \,\, &
\int_0^T \int_{\Omega(t)} f'(\vect{V})[\vect{W}] \dx[V] \dx[t] +
\int_0^T \int_{\partial \Omega(t)}
    f(\vect{V}) \tilde{\vect{V}} \cdot \vect{n} \dx[S] \dx[t], \\
D J_2(\vect{V})[\vect{W}] = \,\, &
\int_0^T \int_{\Sigma(t)} g'(\vect{V})[\vect{W}] \dx[S] \dx[t] +
\int_0^T \int_{\Sigma(t)} (\vect{n} \cdot \nabla g + \kappa g)
    \tilde{\vect{V}} \cdot \vect{n} \dx[S] \dx[t],
\end{aligned}
\end{equation}
where $f'$ and $g'$ denote the shape derivatives of $f$ and $g$, respectively,
as defined in~\cite{Moubachir2006}. In classic shape calculus the perturbation
$\tilde{\vect{V}}$ is taken as a direct perturbation of the geometry itself.
However, for moving domains this perturbation satisfies an evolution equation
of the form
\begin{equation} \label{eq:transverse_field_equation}
\begin{cases}
\tilde{\vect{V}}_t +
    (\vect{V} \cdot \nabla \tilde{\vect{V}} - \tilde{\vect{V}} \cdot \nabla \vect{V})
    = \vect{W},
    & \quad (t, \vect{X}) \in [0, T] \times \mathcal{D}, \\
\tilde{\vect{V}}(0, \vect{x}) = \vect{0},
    & \quad \vect{x} \in \mathcal{D},
\end{cases}
\end{equation}
which ties it to the original perturbation $\vect{W}$ of the velocity field
$\vect{V}$. The equation for the \emph{transverse field} $\tilde{\vect{V}}$
is in fact an ODE if we consider the evolution under the path described by
$\vect{X}(t)$ and driven by the velocity field $\vect{V}$ through~\eqref{eq:evolution}.
For a proof of this fact and additional considerations, we direct the reader
to~\cite{Laurain2021} or~\cite{Moubachir2006}. As the directional derivatives
match those obtained by shape calculus on static domains, we make use of the
results from~\cite{Walker2015} to express any additional shape derivatives for,
e.g., the normal vector, the total curvature, etc.

\subsection{Sensitivity Equations}
\label{ssc:gradient:sensitivity}

A first step in solving the optimal control~\eqref{eq:optim} consists in
determining the sensitivity of the cost functional~\eqref{eq:cost} with
respect to the control. For this, we must consider the cost $J(\vect{g})$ as the
composition of $\vect{g} \mapsto \vect{V}(\vect{g})$ and a shape
functional $\vect{V} \mapsto J(\vect{V})$. For the shape functional, we can
make use of the formulae provided in~\eqref{eq:shape_derivatives}.

Therefore, it remains to find a transverse field equation~\eqref{eq:transverse_field_equation}
that links the perturbations of $\vect{g}$ to the transverse field
$\tilde{\vect{V}}$ and incorporates the constraints given by the two-phase
Stokes equations from~\eqref{eq:stokes}. A similar problem was analyzed
in~\cite{Laurain2021}, where a volume-preserving mean curvature flow was used as
the motion law. Incorporating the two-phase Stokes equations requires a
more careful approach, as we must deal with the discontinuous variables and
jump conditions. To determine the linearized equations, we consider the weak
formulation of the flow, as described in~\cite{Fikl2021}. This weak formulation
contains all the relevant terms and expresses the geometry dependence in an
explicit way, as opposed to the strong form from~\eqref{eq:stokes}. An introduction
to these ideas and applications to simpler elliptic problems are given
in~\cite{Allaire2007}.

We expand on~\cite{Fikl2021} by introducing the tangential field $\vect{w}$
from~\eqref{eq:kinematic} in the adjoint equations. A weak formulation of the
kinematic condition is then given by
\[
\inp{\dot{\vect{X}}, \vect{Y}}_{\Sigma(t)} -
\inp{(\vect{u} \cdot \vect{n}) \vect{n}, \vect{Y}}_{\Sigma(t)}
- \inp{(I - \vect{n} \otimes \vect{n}) \vect{w}, \vect{Y}}_{\Sigma(t)} = 0,
\]
for all test functions $\vect{Y} \in L^2(\Sigma)$. The linearized equations
are then obtained by perturbing the weak formulation using the shape
derivatives from~\eqref{eq:shape_derivatives} and the transverse field
equation~\eqref{eq:transverse_field_equation}. The linearized
two-phase Stokes equations for the shape derivative $(p_\pm', \vect{u}_\pm')$
are given by
\begin{equation} \label{eq:linear:stokes}
\begin{cases}
\nabla \cdot \vect{u}_\pm' = 0,
    & \quad \vect{x} \in \Omega_\pm(t), \\
    \nabla \cdot \sigma_\pm[\vect{u}', p'] = 0,
    & \quad \vect{x} \in \Omega_\pm(t), \\
\vect{u}_+' \to \vect{u}_\infty'[\tilde{\vect{g}}],
    & \quad \|\vect{x}\| \to \infty,
\end{cases}
\end{equation}
with the jump conditions at the interface
\begin{equation} \label{eq:linear:jumps}
\begin{cases}
\jump{\vect{u}'} =
    -\jump{\vect{n} \cdot \nabla \vect{u}} (\tilde{\vect{V}} \cdot \vect{n}), \\
\displaystyle
\jump{\vect{n} \cdot \sigma[\vect{u}', p']}_\lambda =
    \jump{\nabla_\Sigma \cdot
        [(\tilde{\vect{V}} \cdot \vect{n}) \sigma[\vect{u}, p]]}_\lambda
    -\frac{1}{\mathrm{Ca}}
    \Big\{
    (\Delta_\Sigma (\tilde{\vect{V}} \cdot \vect{n})
    + \|\nabla_\Sigma \vect{n}\|^2 \tilde{\vect{V}} \cdot \vect{n}) \vect{n}
    + \kappa \nabla_\Sigma (\tilde{\vect{V}} \cdot \vect{n})
    \Big\}.
\end{cases}
\end{equation}

\begin{remark}
We can see here that the shape derivative $\vect{u}_\pm'$ is not continuous
across the interface like the original velocity field $\vect{u}_\pm$. This
has important consequences and is the reason why we expressed both the
strong form~\eqref{eq:stokes} and the variational formulation on the domains
$\Omega_\pm$ separately, instead of using a \emph{one-fluid}-type formulation.
In particular, even though $\vect{u} \in H^1(\Omega)$, the linear solution
$\vect{u}' \notin H^1(\Omega)$, which implies that the optimization problem
must be posed on $H^1(\Omega_+) \otimes H^1(\Omega_-)$, unlike classic
elliptic shape optimization problems. See~\cite{Pantz2005} for a similar
application to the heat equation with discontinuous coefficients
or~\cite{Feppon2019} for a fluid-structure interaction problem.
\end{remark}

The evolution of the transverse field $\tilde{\vect{V}}$ present in the
above equation is described by
\begin{equation} \label{eq:linear:transverse_field_equation}
\begin{cases}
\pd{\tilde{\vect{V}}}{t} +
    \vect{V} \cdot \nabla \tilde{\vect{V}}
    + \mathcal{K}[\tilde{\vect{V}}]
= \avg*{\vect{u}' \cdot \vect{n}} \vect{n} +
(I - \vect{n} \otimes \vect{n}) \vect{w}',
    & \quad (t, \vect{x}) \in [0, T] \times \Sigma(t), \\
\tilde{\vect{V}}(0, \vect{x}) = 0,
    & \quad \vect{x} \in \Sigma(0),
\end{cases}
\end{equation}
where $\vect{V}$ is prescribed in the kinematic condition~\eqref{eq:kinematic}
and
\[
\mathcal{K}[\tilde{\vect{V}}] \triangleq
- \tilde{\vect{V}} \cdot \nabla \vect{V}
+ \Big\{
\nabla_\Sigma (\tilde{\vect{V}} \cdot \vect{n}) \otimes \vect{n}
+ \vect{n} \otimes \nabla_\Sigma (\tilde{\vect{V}} \cdot \vect{n})
\Big\} (\vect{u} - \vect{w}).
\]

\begin{remark}
While not immediately obvious, the transverse field equation for the two-phase
Stokes system only depends on $\tilde{\vect{V}}$ on $\Sigma(t)$. This is
clear by inspection for the source term $\mathcal{K}[\tilde{\vect{V}}]$. However,
the material derivative
\[
\od{\tilde{\vect{V}}}{t} =
    \pd{\tilde{\vect{V}}}{t} + \vect{V} \cdot \nabla \tilde{\vect{V}}
\]
is also uniquely defined on the interface $\Sigma(t)$, even though its
individual summands are not.

% TODO: The reason it's unique escapes me at the moment. Would be nice to
% $find a reference. \cite{Laurain2021} also just states it matter-of-fact-ly.
\end{remark}

We could now use the shape derivatives from~\eqref{eq:shape_derivatives} to
perturb the cost functional~\eqref{eq:cost} and obtain an expression in
terms of the sensitivities $(p_\pm', \vect{u}_\pm', \tilde{\vect{V}})$.
However, it is well-known that this method is prohibitively costly when
performing numerical experiments, as it requires many solutions
to~\eqref{eq:linear:stokes} and~\eqref{eq:linear:transverse_field_equation}. A
significant saving can be obtained by expressing the gradient in terms of
adjoint variables.

\subsection{Adjoint Equations}
\label{ssc:gradient:adjoint}

The adjoint problem can be obtained directly from the weak formulation of the
sensitivity equations from~\Cref{ssc:gradient:sensitivity} by applying the
appropriate integration by parts theorems. This derivation is provided in
detail in~\cite{Fikl2021} through a Lagrangian formalism. The adjoint
two-phase Stokes equations are given by
\begin{equation} \label{eq:adjoint:stokes}
\begin{cases}
\nabla \cdot \vect{u}^*_\pm = 0,
    & \quad \vect{x} \in \Omega_\pm(t), \\
\nabla \cdot \sigma_\pm[\vect{u}^*, p^*] = 0,
    & \quad \vect{x} \in \Omega_\pm(t), \\
\vect{u}_+^* \to 0,
    & \quad \|\vect{x}\| \to \infty,
\end{cases}
\end{equation}
with the jump conditions
\begin{equation} \label{eq:adjoint:jumps}
\begin{cases}
\jump{\vect{u}^*} = 0, \\
\jump{\vect{n} \cdot \sigma[\vect{u}^*, p^*]}_\lambda =
    (\vect{X}^* \cdot \vect{n}) \vect{n}
    + (\vect{u} \cdot \vect{n} - u_d) \vect{n},
\end{cases}
\end{equation}
where $(p_\pm^*, \vect{u}_\pm^*)$ are the adjoint pressure and velocity fields.
Similarly, we obtain a backwards evolution equation for an adjoint transverse
field variable $\vect{X}^*$. It is given by
\begin{equation} \label{eq:adjoint_transverse_field}
\begin{cases}
\displaystyle
-\od{\vect{X}^*}{t}
    - \vect{X}^* \cdot \nabla \vect{V}^T
    - \vect{X}^* \nabla_\Sigma \cdot \vect{V} =
    \left\{
    j^*_\Sigma[\vect{v}, \vect{X}]
    + \mathcal{K}^*[\vect{X}^*; \vect{v}, \vect{X}]
    + \mathcal{S}^*[\vect{v}^*; \vect{v}, \vect{X}]
    + \frac{1}{\mathrm{Ca}} \mathcal{C}^*[\vect{v}^*; \vect{v}, \vect{X}]
    \right\} \vect{n}, \\
\vect{X}^*(T) = j^*_{\Sigma_T}[\vect{v}, \vect{X}] \vect{n},
\end{cases}
\end{equation}
where the source terms are given by
\begin{equation} \label{eq:adjoint_transverse_field_sources}
\begin{aligned}
j^*_\Sigma[\vect{v}, \vect{X}] \triangleq \,\, &
    \alpha_1 \left\{
    \nabla_\Sigma \cdot [j_u \vect{u}]
    - \kappa j_u (\vect{u} \cdot \vect{n})
    + j_u \avg{\vect{n} \cdot \nabla \vect{u}} \cdot \vect{n}
    + \frac{1}{2} \kappa j_u^2
    \right\}
    + \frac{\alpha_2}{|\Omega_-|}
    (\vect{x}_c - \vect{x}_d) \cdot (\vect{X} - \vect{x}_c),
    \\
j^*_{\Sigma_T}[\vect{v}, \vect{X}] \triangleq \,\, &
    \frac{\alpha_3 }{|\Omega_-|}
    (\vect{x}_c(T) - \vect{x}_{d, T}) \cdot (\vect{X}(T) - \vect{x}_c(T)), \\
\mathcal{K}^*[\vect{X}^*; \vect{v}, \vect{X}] \triangleq \,\, &
\nabla_\Sigma \cdot \left[
    ((\vect{u} - \vect{w}) \cdot \vect{n}) \vect{X}^*
    + (\vect{X}^* \cdot \vect{n}) (\vect{u} - \vect{w})
    \right]
    - 2 \kappa ((\vect{u} - \vect{w}) \cdot \vect{n}) (\vect{X}^* \cdot \vect{n}), \\
\mathcal{S}^*[\vect{v}^*; \vect{v}, \vect{X}] \triangleq \,\, &
    2 \jump{\varepsilon[\vect{u}] \cdot \varepsilon[\vect{u}^*]}_\lambda
    - \avg{\vect{n} \cdot \sigma}_\lambda \cdot \jump{\vect{n} \cdot \nabla \vect{u}^*}
    - \avg{\vect{n} \cdot \sigma^*}_\lambda \cdot \jump{\vect{n} \cdot \nabla \vect{u}}, \\
\mathcal{C}^*[\vect{v}^*; \vect{v}, \vect{X}] \triangleq \,\, &
    \Delta_\Sigma \avg{\vect{u}^* \cdot \vect{n}}
    + \|\nabla_\Sigma \vect{n}\|^2 \avg{\vect{u}^* \cdot \vect{n}}
    - \nabla_\Sigma \cdot (\kappa \vect{u}^*)
    - \kappa \avg{\vect{n} \cdot \nabla \vect{u}^*} \cdot \vect{n},
\end{aligned}
\end{equation}
where $j_u \triangleq \vect{u} \cdot \vect{n} - u_d$ and
$\vect{v} \triangleq (p_\pm, \vect{u}_\pm)$ and $\vect{v}^* \triangleq
(p_\pm^*, \vect{u}_\pm^*)$ denote the coupled state and adjoint variables,
respectively. We have grouped the terms arising from the kinematic
condition~\eqref{eq:kinematic} in $\mathcal{K}^*$, those arising from the
Stokes system~\eqref{eq:stokes} in $\mathcal{S}^*$, and those arising from the
traction jump conditions~\eqref{eq:jumps} in $\mathcal{C}^*$.

\begin{remark}
As the sensitivity equations from~\Cref{ssc:gradient:sensitivity}, the
adjoint transverse field equations are well-defined on the interface $\Sigma(t)$.
The only terms of interest involve the state and adjoint velocity fields.
However, we know that
\[
\jump{\vect{u} \cdot \vect{n}}
    = \jump{(\vect{n} \cdot \nabla \vect{u}) \cdot \vect{n}}
    = 0,
\]
for incompressible flows with no phase change.
\end{remark}

In~\eqref{eq:adjoint_transverse_field_sources}, we have given the source
terms as obtained by directly applying~\eqref{eq:shape_derivatives} and
integration by parts. However, several simplifications are possible.
First, we have that
\begin{equation} \label{eq:adjoint_source_c}
\begin{aligned}
\mathcal{C}^*[\vect{v}^*; \vect{v}, \vect{X}] =\,\, &
\vect{n} \cdot \Delta \vect{u}^*
- \vect{n} \cdot (\vect{n} \cdot \nabla \nabla \vect{u}^*) \vect{n}
+ 2 \nabla \vect{u}^* \cdot \nabla_\Sigma \vect{n}
- \kappa (\vect{n} \cdot \nabla \vect{u}^*) \cdot \vect{n},
\end{aligned}
\end{equation}
where we have expanded the Laplace-Beltrami operator using the standard
expression
\begin{equation} \label{eq:adjoint_source_s}
\Delta_\Sigma f \triangleq
\Delta f - \kappa \vect{n} \cdot \nabla f - \vect{n} \cdot \nabla \nabla f \cdot \vect{n}
\end{equation}
and performed the required simplifications. We can see that, in this form,
we do not require third-order derivatives of the geometry, e.g. in the
$\nabla_\Sigma \cdot (\kappa \vect{u}^*)$ term. Similarly, it is inconvenient
to express the full rate-of-strain tensors $\varepsilon_\pm$ in the source term.
We can decompose them into the orthonormal basis $(\vect{n}, \vect{t}^\alpha)$,
where $\vect{t}^\alpha$ is an orthonormal basis of tangent space at every
point on $\Sigma(t)$. This gives
\[
\begin{aligned}
\mathcal{S}^*[\vect{v}^*; \vect{v}, \vect{X}] =\,\, &
\jump{\vect{n} \cdot \sigma^*}_\lambda \cdot \avg{\vect{n} \cdot \nabla \vect{u}}
+ \avg{\vect{t}^\alpha \cdot \nabla \vect{u}}
    \cdot \jump{\vect{t}^\alpha \cdot \sigma^*}_\lambda
- \avg{\vect{n} \cdot \sigma}_\lambda \cdot \jump{\vect{n} \cdot \nabla \vect{u}^*} \\
=\,\, &
\jump{\vect{n} \cdot \sigma}_\lambda \cdot \avg{\vect{n} \cdot \nabla \vect{u}^*}
+ \avg{\vect{t}^\alpha \cdot \nabla \vect{u}^*}
    \cdot \jump{\vect{t}^\alpha \cdot \sigma}_\lambda
- \avg{\vect{n} \cdot \sigma^*}_\lambda \cdot \jump{\vect{n} \cdot \nabla \vect{u}},
\end{aligned}
\]
where summation over the repeated $\alpha$ indices is implied.
In the case $\lambda = 1$, we also have $\jump{\vect{n} \cdot \nabla \vect{u}}
= \jump{\vect{n} \cdot \nabla \vect{u}^*} = 0$, which allows further terms to
cancel. Similarly, if the surface traction jumps have no tangential
components, the corresponding terms vanish. These simplifications are used,
as appropriate, in the numerical results from~\Cref{sc:results}. Finally,
we are in a position to express the adjoint-based gradient of the cost
functional from~\eqref{eq:cost} with respect to the full farfield velocity field
$\vect{u}_\infty$. Expressions for the gradient with respect to a specific
parameter $\vect{g}$ used to describe $\vect{u}_\infty$ can be obtained by the
chain rule. The directional derivative is given by
\[
\begin{aligned}
D J[\tilde{\vect{u}}_\infty] =\,\, &
\alpha_1 \int_0^T \int_{\Sigma(t)}
    (\vect{u} \cdot \vect{n} - u_d) \vect{n} \cdot \tilde{\vect{u}}_\infty
    \dx[S] \dx[t] \\
+\,\, & \int_0^T \int_{\Sigma(t)}
(\vect{X}^* \cdot \vect{n}) \vect{n} \cdot \tilde{\vect{u}}_\infty \dx[S] \dx[t]
+ (1 - \lambda) \int_0^T \int_{\Sigma(t)}
\vect{u}^* \cdot (\vect{n} \cdot \varepsilon[\tilde{\vect{u}}_\infty])
\dx[S] \dx[t],
\end{aligned}
\]
from which we can extract the standard $L^2$ gradient by the Riesz Representation
theorem
\begin{equation} \label{eq:adjoint_gradient}
\nabla_{\vect{u}_\infty} J =
\Big\{
\alpha_1 (\vect{u} \cdot \vect{n} - u_d)
+ \vect{X}^* \cdot \vect{n}
+ (1 - \lambda) \varepsilon^*[\vect{u}^* \otimes \vect{n}]
\Big\} \vect{n}.
\end{equation}

\subsection{Normal Transverse Field Evolution Equations}
\label{ssc:gradient:normal}

In~\Cref{ssc:gradient:sensitivity} and~\Cref{ssc:gradient:adjoint} we
derived the evolutions equations for the transverse field $\tilde{\vect{V}}$
and the adjoint transverse field $\vect{X}^*$. However, we notice that the
gradient of the cost functional~\eqref{eq:adjoint_gradient} only depends on the
normal component of the adjoint field. Furthermore, most of the source
terms~\eqref{eq:adjoint_transverse_field_sources} in the adjoint transverse
field equation and the adjoint jump conditions~\eqref{eq:adjoint:jumps} also
only contain the normal component. As such, it is natural to ask whether we
can find an evolution equation only for the normal component
$X^*_n \triangleq \vect{X}^* \cdot \vect{n}$. Numerically, this simplification
is beneficial, as we can solve a simpler scalar equation.

For brevity, we will focus on the adjoint equations, but an evolution equation
for the normal transverse field $\tilde{V}_n \triangleq \tilde{\vect{V}} \cdot
\vect{n}$ can also be obtained in an analogous manner (see~\cite{Laurain2021}).
For the adjoint field, we have that
\begin{equation} \label{eq:normal_adjoint_transverse_field}
\begin{cases}
-\od{X^*_n}{t} - (\vect{u} - \vect{w}) \cdot \nabla_\Sigma X^*_n =
j^*_\Sigma + \mathcal{S}^*[\vect{v}^*; \vect{v}, \vect{X}]
+ \mathcal{C}^*[\vect{v}^*; \vect{v}, \vect{X}], \\
X^*_n(0) = j^*_{\Sigma_T}.
\end{cases}
\end{equation}

We give below a short proof of this derivation, as it is the first large
departure from the results presented in~\cite{Fikl2021}, where the adjoint
transverse field $\vect{X}^*$ was used. A similar proof can be found
in~\cite{Laurain2021} for the aforementioned mean curvature flow case.

\begin{proof}
We proceed by showing that the tangential components of $\vect{X}^*$ vanish
for all times. Then, it suffices to decompose the adjoint transverse field
into its normal and tangential components and dot the transverse field
equation~\eqref{eq:adjoint_transverse_field} with the normal vector to
obtain~\eqref{eq:normal_adjoint_transverse_field}.

We consider the tangential components defined by projection $\vect{X}^*_\tau \triangleq
(I - \vect{n} \otimes \vect{n}) \vect{X}^*$. It is clear
from~\eqref{eq:adjoint_transverse_field} that $\vect{X}^*_\tau(T) = 0$ as the
``initial'' condition. Therefore, it remains to analyze the equation itself.
We start by applying the same projection and remove the terms that only have
a normal component. This leaves
\[
-(I - \vect{n} \otimes \vect{n}) \left[
    \od{\vect{X}^*}{t} + \vect{X}^* \cdot \nabla \vect{V}^T
    + \vect{X}^* \nabla_\Sigma \cdot \vect{V}
    \right] = 0.
\]

To apply the product rule and include the projection into the time derivative,
we have the following expression for the time rate of change of the normal
vector (see~\cite[Lemma 3.3]{Huisken1984} or~\cite[Lemma 5.5]{Walker2015})
\[
\od{\vect{n}}{t} = -\vect{n} \cdot \nabla_\Sigma \vect{V}^T.
\]

Using this expression, we have that the projection of the time rate-of-change
becomes
\[
\begin{aligned}
(I - \vect{n} \otimes \vect{n}) \od{\vect{X}^*}{t}
=\,\, & \od{\vect{X}^*_\tau}{t} + \left[
    \od{\vect{n}}{t} \otimes \vect{n} + \vect{n} \otimes \od{\vect{n}}{t}
    \right] \vect{X}^* \\
=\,\, &
\od{\vect{X}^*_\tau}{t} -
(\vect{X}^* \cdot \vect{n}) (\vect{n} \cdot \nabla_\Sigma \vect{V}^T)
- (\vect{n} \cdot (\vect{X}^*_\tau \cdot \nabla_\Sigma \vect{V})) \vect{n}.
\end{aligned}
\]

For the remaining source terms, we have that
\[
\begin{aligned}
(I - \vect{n} \otimes \vect{n})
    (\vect{X}^* \cdot \nabla \vect{V}^T + \vect{X}^* \nabla_\Sigma \cdot \vect{V})
=\,\, & \vect{X}^* \cdot \nabla_\Sigma \vect{V}^T
    + \vect{X}^*_\tau \nabla_\Sigma \cdot \vect{V} \\
=\,\, &
    \vect{X}^*_\tau \cdot \nabla_\Sigma \vect{V}^T
    + (\vect{X}^* \cdot \vect{n}) (\vect{n} \cdot \nabla_\Sigma \vect{V}^T)
    + \vect{X}^*_\tau \nabla_\Sigma \cdot \vect{V},
\end{aligned}
\]
where we can see that $\vect{X}^*_\tau$ does not appear in any derivatives.
Putting the two parts back together, we can see that the $\vect{X}^* \cdot \vect{n}$
term cancels out and we are left with an equation of the form
\[
-\od{\vect{X}^*_\tau}{t} = \vect{X}^*_\tau \cdot \Big\{
    \nabla_\Sigma \vect{V}^T
    + \nabla_\Sigma \cdot \vect{V} I
    - \nabla_\Sigma \vect{V} (\vect{n} \otimes \vect{n})
    \Big\},
\]
where the right-hand side does not contain any source terms that do not
depend on $\vect{X}^*_\tau$ linearly. As $\vect{X}^*_\tau(T) = 0$, the above
Lagrangian transport will maintain this ``initial'' value, so
$\vect{X}^*_\tau(t) \equiv 0$ for all times $t \in [0, T]$.
\end{proof}

\begin{remark}
Note that only the tangential components of $\vect{w}$ are required in
defining the state and the adjoint systems. This is not obvious in the
adjoint transverse field equation~\eqref{eq:adjoint_transverse_field}, but can be
clearly seen in the scalar equation~\eqref{eq:normal_adjoint_transverse_field}.
\end{remark}

\section{Numerical Methods}
\label{sc:methods}

We present here a discretization of the quasi-static two-phase Stokes equations
from~\eqref{eq:stokes}. In our discretization, we assume that no topological
changes occur (such as breakup) and all variables maintain sufficient
regularity. This is not the case in practice and we will present a simple and
effective filtering method. Finally, we use the optimize-then-discretize path
to adjoint-based optimization, so the discrete state and adjoint problems
will only be consistent in the limit of vanishing grid sizes.
See~\cite{Gunzburger2003} for a discussion on the pitfalls involved in this
choice, as opposed to the discretize-then-optimize approach.

As we have seen in the previous section, we require high-order derivatives
of both the geometry and the forward and adjoint variables. As expected, these
issues mainly come into play in the adjoint problem and, in particular, in the
adjoint transverse equation~\eqref{eq:normal_adjoint_transverse_field}.
For the geometry, we will be using a representation based on spherical
harmonics, as described, e.g., in~\cite{Veerapaneni2011}. Then, for the
Stokes solver, we apply the popular methods based on boundary integral
equations (see~\cite{Pozrikidis1992}). The main difficulty in the boundary
integral equation-based methods is expressing the singularities numerically
in an accurate way. To ensure that we have a robust solver, we make use of the
QBX method~\cite{Klockner2013}. This method allows handling arbitrary
singularities in a kernel-independent way, which is very beneficial for our
problem, as we require a large number of kernels to express all the terms
in~\eqref{eq:normal_adjoint_transverse_field}. We will go into detail on
these methods in the following sections.

\subsection{Nyström-QBX Boundary Integral Discretization}
\label{ssc:methods:qbx}

To solve the two-phase Stokes equations~\ref{eq:stokes}, we will make use of the
single-layer representation presented in~\cite{Fikl2021}. In this formulation,
the pressure and the velocity field are represented by
\begin{equation} \label{eq:velocity}
\begin{aligned}
u_{\pm, i}(\vect{x}) = \,\, & u_{\infty, i}(\vect{x}) +
\int_\Sigma G_{ij}(\vect{x}, \vect{y}) q_j(\vect{y}) \dx[S], \\
p_{\pm, i}(\vect{x}) =\,\, & p_{\infty}(\vect{x}) +
\int_{\Sigma} P_{j}(\vect{x}, \vect{y}) q_j(\vect{y}) \dx[S],
\end{aligned}
\end{equation}
for all $\vect{x} \notin \Sigma$. The usual summation conventions
for repeated indices are used. The kernels in the two integrals are given by
\[
\begin{aligned}
G_{ij}(\vect{r}) \triangleq \,\, & -\frac{1}{8 \pi}
\left(\frac{1}{r} \delta_{ij} + \frac{r_i r_j}{r^3}\right), \\
P(\vect{r}) \triangleq \,\, & -\frac{1}{4 \pi} \frac{r_j}{r^3},
\end{aligned}
\]
where $\vect{r} \triangleq \vect{x} - \vect{y}$ and $r \triangleq \|\vect{r}\|$.
To solve the boundary integral equation and obtain the solutions to Stokes
flow, we must find the density $\vect{q}(\vect{x})$. For a two-phase flow,
we make use of the jump in surface traction and write
\begin{equation} \label{eq:density}
\jump{n_i \sigma_{ik}}_\lambda =
(1 - \lambda) n_i \sigma_{\infty, ik}(\vect{x})
+ \frac{1 + \lambda}{2} q_k(\vect{x})
+ (1 - \lambda)~ \mathrm{p.v.} \int_{\Sigma}
    n_i(\vect{x}) T_{ijk}(\vect{x}, \vect{y}) q_j(\vect{y}) \dx[S],
\end{equation}
where $\mathrm{p.v.}$ denotes the Cauchy Principal Value interpretation of the
integral. Here, $T_{ijk}$ is known as the Stresslet and is given by
\[
T_{ijk}(\vect{r}) \triangleq \frac{1}{4 \pi} \frac{r_i r_j r_k}{r^5}.
\]

The boundary integral equation given in~\eqref{eq:density} is a Fredholm
integral equation of the second kind. As such, we know from~\cite{Pozrikidis1992}
that it has solutions for all right-hand sides and the operator in question
is well-conditioned. In the special case $\lambda = 1$, the density can
be determined directly from the surface traction boundary conditions, without
the need to solve a system. For the state equations, it suffices to find the
velocity field to evolve the interface using the kinematic condition~\eqref{eq:kinematic}.
However, for the normal adjoint transverse equation~\eqref{eq:normal_adjoint_transverse_field}
we require additional expressions for the components of the velocity
gradient, tangential components of the stress and others. These layer potentials
and their corresponding jump conditions are provided in~\cite{Fikl2021}.
For this work, we also require an expression for the velocity Laplacian and
Hessian that appear in~\eqref{eq:adjoint_source_c}. They are given by
\[
\begin{aligned}
n_i \pd{u_{\pm, i}}{x_k \partial x_k} =\,\, &
n_i \pd{u_{\infty, i}}{x_k \partial x_k}
+ \int_{\Sigma} n_i(\vect{x})
    \pd{G_{ij}}{x_k \partial x_k}(\vect{x}, \vect{y}) q_j(\vect{y}) \dx[S], \\
n_i n_k n_l \pd{u_{\pm, i}}{x_k \partial x_l} =\,\, &
n_i n_k n_l \pd{u_{\infty, i}}{x_k \partial x_l}
+ \int_{\Sigma} n_i(\vect{x}) n_k(\vect{x}) n_l(\vect{x})
    \pd{G_{ij}}{x_k \partial x_l}(\vect{x}, \vect{y}) q_j(\vect{y}) \dx[S], \\
\end{aligned}
\]
for all $\vect{x} \notin \Sigma$. The surface limits of these quantities
lead to hypersingular integrals that must be interpreted through the Hadamard
Finite Part regularization. Their surface limits are given in~\cite{Fikl2020}.

As we have seen, to fully express the source terms in the adjoint transverse
field equation~\eqref{eq:normal_adjoint_transverse_field}, we are required to
evaluate integrals with many different types of singularities. Here, the
velocity field is weakly singular, the traction and velocity gradients are
strongly singular, while the Laplacian and Hessian are hypersingular.
Many standard methods in the literature require special treatment for each
kind of singularity and kernel, which is prohibitively complex for our
problem. We have therefore chosen to use the QBX method~\cite{Klockner2013},
as a regularization procedure inside a standard Nyström method~\cite{Kress1989},
that can work across different types of kernels and singularities with
predictable accuracy. We make use of its implementation in the
\texttt{pytential}~\cite{pytential} open-source library and extend it
to express all the layer potentials we have discussed above. The implementation
makes use of a state-of-the-art FMM (Fast Multipole Method) tailored for
QBX~\cite{Wala2019}. The Stokes kernels are expressed using~\cite{Tornberg2008}
for improved efficiency.

\begin{figure}[ht!]
\centering
\includegraphics[trim=700 350 700 350, clip, width=0.55\linewidth]{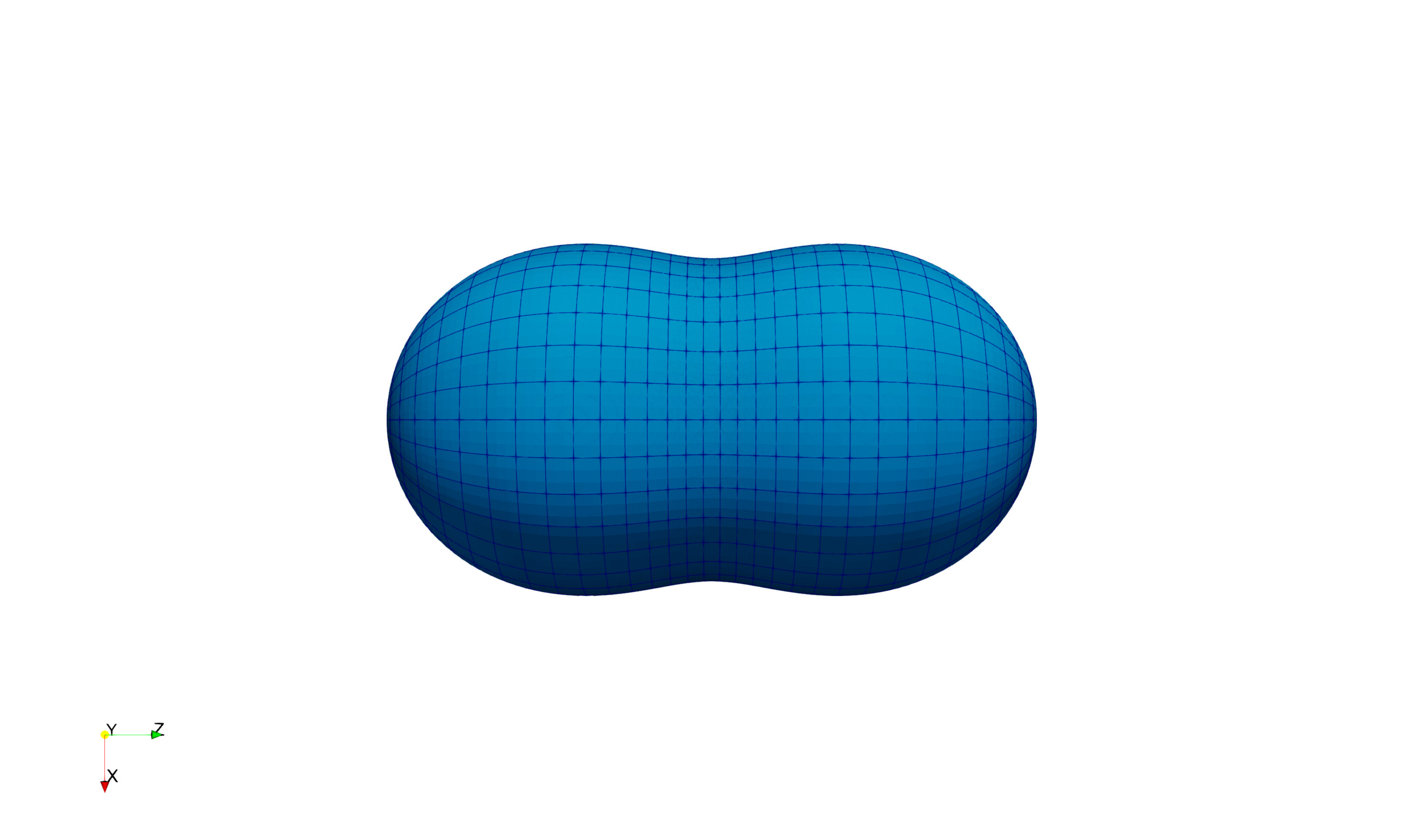}
\caption{Surface discretization based on quadrilateral elements.}
\label{fig:discretization}
\end{figure}

To fix some nomenclature for the following sections, we briefly describe
the parameters required to define the discretization (see~\Cref{fig:discretization}).
First, the surface $\Sigma$ is discretized by quadrilateral elements, where
the reference element is given by $[-1, 1] \times [-1, 1]$. On these elements,
tensor product Gauss-Legendre nodes and weights of order $P$ are used to provide
the quadrature rule for the Nyström method. In-element interpolation is performed
using the orthonormal Legendre polynomials of the same order. For the spherical
harmonic representation of~\Cref{ssc:spharm}, we require an equidistant
discretization in the $(\theta, \phi)$ charts. For this, we use $M_\theta$
and $M_\phi$ elements in the $\theta$ and $\phi$ directions, respectively,
for a total of $M \triangleq M_\theta \times M_\phi$ elements. Then,
the QBX method requires a local expansion of the kernel up to order
$P_{qbx}$. The QBX method, as implemented in \texttt{pytential}, uses an
oversampled quadrature grid of order $P_{quad}$ (on top of the base grid of
order $P$) to ensure sufficient resolution (see~\cite{Wala2019} for
additional details). Finally, for the FMM, additional expansions (local and
multipole) of order $P_{fmm}$ are required.

\subsection{Spherical Harmonics and Filtering}
\label{ssc:spharm}

\begin{figure}[ht!]
\centering
\begin{tikzpicture}
\begin{scope}[xshift=3.5cm, yshift=0.25cm]
\draw[domain=-0.5:1.6, samples=128, color=JCPOrange] plot (\x, {0.5+0.25*((sin(360*\x)+cos(2*360*\x)+cos(1*360*\x))});
\draw[domain=-0.5:1.6, samples=128, color=JCPLightBlue] plot (\x, {0.5+0.25*((-cos(360*\x)+cos(2*360*\x)+sin(3*360*\x))});
\end{scope}

\begin{scope}[xshift=-2.25cm, yshift=0.25cm]
\draw[thick] (0, 0) rectangle (1, 1) rectangle (2, 0);
\foreach \y in {0, 1} {
    \foreach \x in {0, 1.0, 2.0} {
        \fill[JCPOrange] (\x, \y) circle [radius=0.08];
    }
}
\foreach \y in {0, 0.5, 1} {
    \foreach \x in {0, 0.5, 1.0, 1.5, 2.0} {
        \fill[black] (\x, \y) circle [radius=0.04];
    }
}
\end{scope}

\begin{scope}[xshift=-2cm, yshift=-2.5cm]
\draw[thick] (-0.5, 0) rectangle (0.5, 1);
\foreach \x in {-0.5, 0.0, 0.5} {
    \foreach \y in {0, 0.5, 1} {
        \fill[JCPLightBlue] (\x, \y) circle [radius=0.04];
    }
}

\draw[thick] (1, 0) rectangle (2, 1);
\foreach \y in {0, 0.5, 1} {
    \foreach \x in {1, 1.5, 2} {
        \fill[JCPLightBlue] (\x, \y) circle [radius=0.04];
    }

    \draw[black!40, <->] (0.6, \y) -- (0.9, \y);
}
\end{scope}

\begin{scope}[xshift=3.5cm, yshift=-2.5cm]
\draw[thick] (-0.5, 0) rectangle (0.5, 1);
\foreach \x in {-0.38729833, 0.0, 0.38729833} {
    \foreach \y in {0.11270167, 0.5, 0.88729833} {
        \fill[JCPLightBlue] (\x, \y) circle [radius=0.04];
    }
}

\draw[thick] (0.6, 0) rectangle (1.6, 1);
\foreach \y in {0.11270167, 0.5, 0.88729833} {
    \foreach \x in {0.71270167, 1.1       , 1.48729833} {
        \fill[JCPLightBlue] (\x, \y) circle [radius=0.04];
    }
}
\end{scope}

\draw[thick, ->] (4.05, 1.5) to [in=15, out=165] (-1.25, 1.5);
\draw[thick, <-, dashed] (4.05, 0.0) to [in=-15, out=-165] (-1.25, 0.0);

\draw[thick, ->] (-1.75, 0)  --(-1.75, -1.3);
\draw[thick, dashed, <-] (-1.5, 0.0) -- (-1.5, -1.3);

\draw[thick, ->] (-1.25, -2.75) to [in=-165, out=-15] (4.05, -2.75);
\draw[thick, dashed, <-] (-1.25, -1.25) to [in=165, out=15] (4.05, -1.25);

\node at (5.25, 0.75) [right] {\textbf{Layer 1}};
\node at (-2.5, 0.75) [left] {\textbf{Layer 2}};
\node at (1.4, -2.0) {\textbf{Layer 3}};
\end{tikzpicture}

\caption{
    Projection to and from the spherical harmonic coefficients and the
    QBX quadrature grid. (a) Layer 1 represents the spectral coefficients, (b)
    Layer 2 represents an equidistant grid used by the \texttt{SHTns}~\cite{SHTns}
    library to compute the coefficients, and (c) Layer 3 represents the
    discontinuous grid of quadrature nodes used by the Nyström-QBX method.}
\label{fig:layers}
\end{figure}
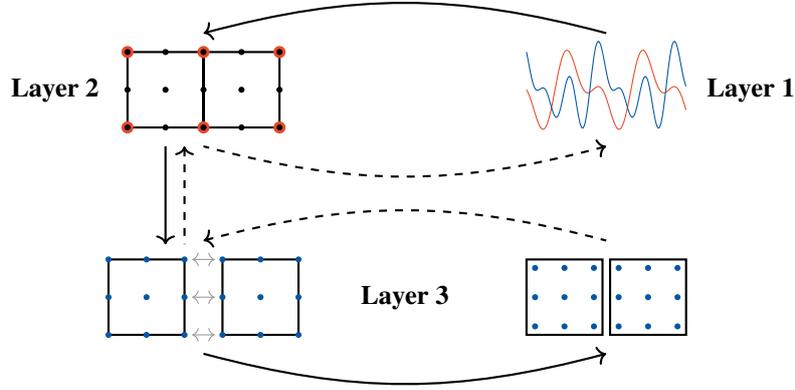

We use spherical harmonics to represent the surface and all related geometric
quantities, such as the normal and curvature. For a similar application to
Stokes flow, see~\cite{Veerapaneni2011}, where spherical harmonics were also
used to express the layer potentials. The spherical harmonic transforms are
performed with the aid of the \texttt{SHTns}~\cite{SHTns} open-source library
described in~\cite{Schaeffer2013}. For a  smooth closed surface (with a
spherical topology) $\Sigma$ and a chosen parametrization
$\vect{X}(\theta, \phi): [0, \pi] \times [0, 2 \pi] \to \Sigma$, we
express
\[
\vect{X}(\theta, \phi) = \sum_{n = 0}^{N_\theta} \sum_{m = 0}^{\min(n, N_\phi)}
    \vect{X}^m_n Y^m_n(\theta, \phi),
\]
where $Y^m_n$ are the spherical harmonics of order $m$ and degree $n$. They
are given by
\[
Y^m_n(\theta, \phi) \triangleq
\sqrt{\frac{2 n + 1}{4} \frac{(n - m)!}{(n + m)!}} P^m_n(\cos \theta) e^{\imath m \phi}
\quad \text{and} \quad
P^m_n(x) \triangleq (-1)^m \frac{(1 - x^2)^{\frac{m}{2}}}{2^n n!}
\od{^{m + n}}{x^{m + n}} (x^2 - 1)^n,
\]
where $P^m_n(x)$ are the associated Legendre functions of the first kind
with the Cordon-Shortley phase included as a matter of convention. The
spherical harmonics defined as above form an orthonormal basis for square-integrable
functions on the unit sphere $\mathbb{S}^2$ when $N_\theta, N_\phi \to \infty$.
In practice, we choose
\[
N_\theta \lesssim \left\lfloor \frac{P M_\theta}{2} \right\rfloor
\quad \text{and} \quad
N_\phi \lesssim \left\lfloor \frac{P M_\phi}{2} \right\rfloor,
\]
as required by the \texttt{SHTns} library. The main difficulty in our
discretization arrives when attempting to project between the nodes used
by the QBX method and the spherical harmonic coefficients. We perform this
projection in three steps, as shown in~\Cref{fig:layers}.

First, the \texttt{SHTns}~\cite{SHTns} library uses a uniform grid, for which
FFT-type methods are well-suited and provide a substantial speed-up. On the
other hand, the Nyström-QBX method described in~\Cref{ssc:methods:qbx}
is implemented on a Discontinuous Galerkin-type quadrature grid, where
continuity between the elements is not enforced. The roundtrip is performed
as follows:
\begin{itemize}
    \item \textbf{Backward}. From the spectral coefficients, we use the
    \texttt{SHTns} library to obtain the physical values on the equidistant
    Layer 2 grid. These values are then copied to an equidistant grid
    where each node is unique. Finally, the values are interpolated element-wise
    to the Layer 3 quadrature grid using the Legendre polynomials.
    \item \textbf{Forward}. From the quadrature nodes on the Layer 3 grid, we
    interpolate back to an equidistant grid on the reference element. Then, to
    obtain unique values at vertices and faces, the repeated values on the
    element faces are averaged. Finally, the \texttt{SHTns} library is used to
    obtain the spectral coefficients.
\end{itemize}

\begin{figure}[ht!]
\begin{subfigure}[b]{0.5\linewidth}
\centering
\includegraphics[width=0.8\linewidth]{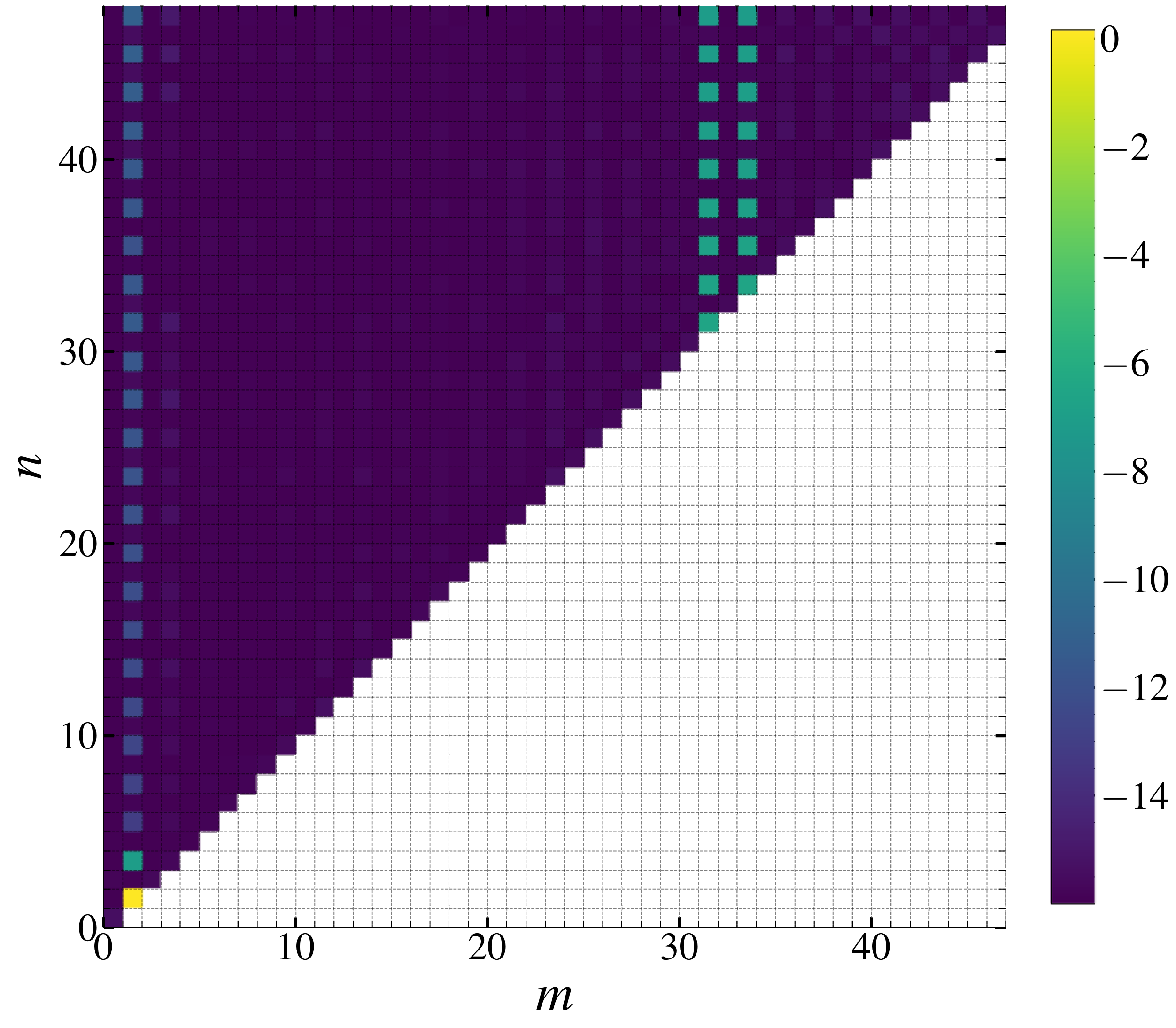}
\caption{}
\end{subfigure}
\hfill
\begin{subfigure}[b]{0.5\linewidth}
\centering
\includegraphics[width=0.8\linewidth]{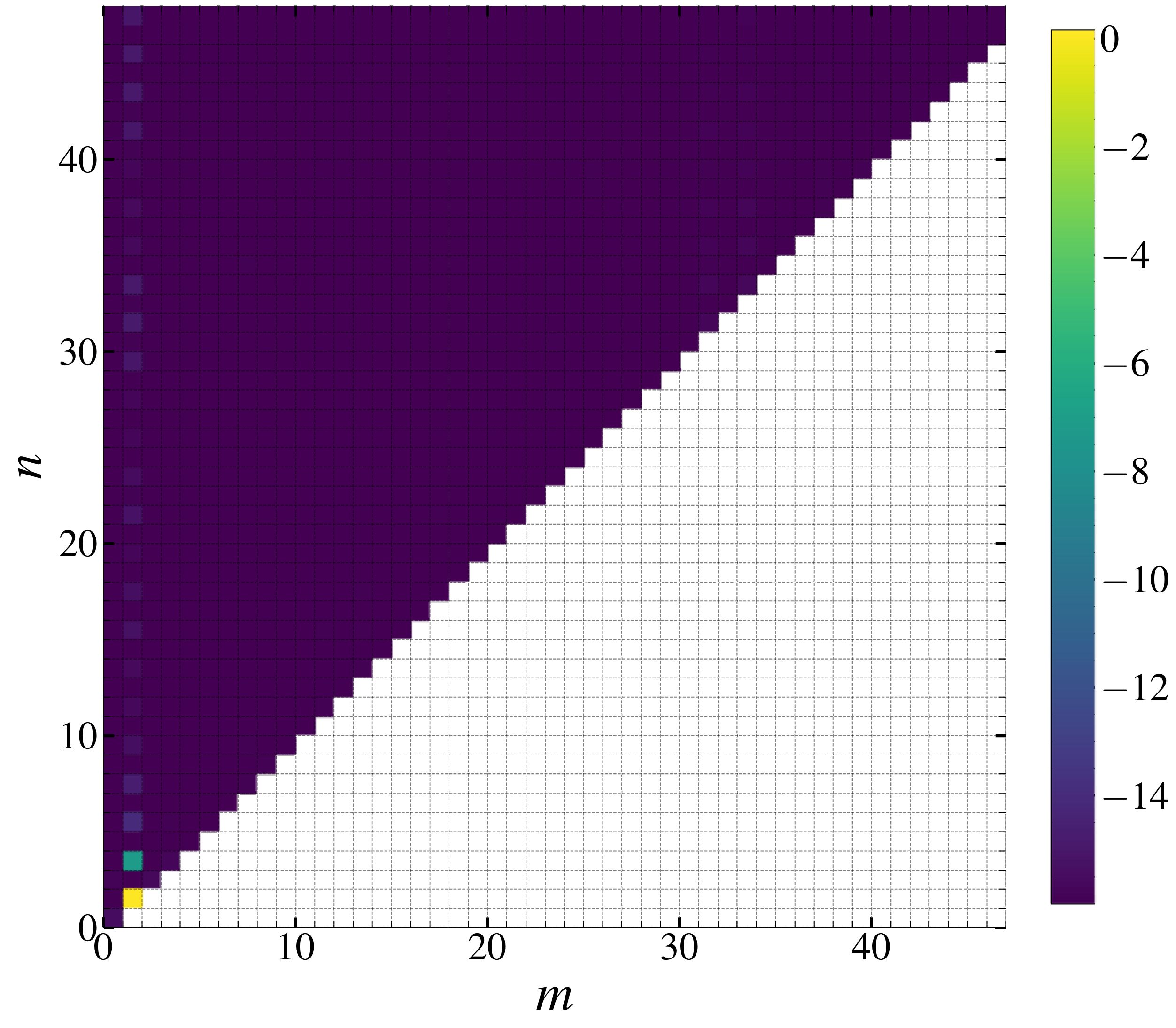}
\caption{}
\end{subfigure}
\caption{
    Filtering: Spherical harmonic coefficients of the $x$ component of the
    normal vector on a unit sphere. (a) Original projection and (b)
    filtered projection by~\eqref{eq:filter} with $\alpha = 10^{-5}$ and
    $p = 1$. The mesh used $M_\theta = M_\phi = 32$ elements and $P = 3$.}
\label{fig:normal}
\end{figure}

As we can see, there are several sources of unwanted errors in this construction.
First, in both the forward and backward transforms, we have an interpolation
from the chosen quadrature grid. This operation is well-conditioned, due to the
choice of Gauss-Legendre nodes and the Legendre orthonormal polynomials on the
reference element. The main source of error comes from the averaging performed
in the forward transform to obtain unique values at the vertices and faces of
each element. This operation can introduce spurious high-frequency components
into the spectrum and rapidly lead to growing instabilities in the evolution
of the droplets. An example of the resulting errors can be seen in~\Cref{fig:normal},
where the normal vector was computed on the unit sphere and then projected
to its spherical harmonic coefficients. We know that the $x$ component of the
normal vector only contains the $Y^1_1(\theta, \phi)$ mode, but additional
modes due to the interpolation and averaging appear in the result.

In general, we know that the errors introduced by the method must have a
frequency that is proportional to the element grid spacing in each direction
$(\theta, \phi)$. Therefore, a method that can alleviate these problems
would be to simply filter out all modes above a certain wavenumber using a
standard ideal filter. This is possible in the $\phi$ direction, where the
modes are essentially Fourier modes. However, in the $\theta$ direction, we
encounter the associated Legendre functions, where the unique Fourier modes
cannot be obtained, and filtering is not as clear. For these reasons, we
adopt a hybrid scheme with the filter
\begin{equation} \label{eq:filter}
\hat{F}^m_n =
\begin{cases}
\displaystyle
\frac{F^m_n}{1 + \alpha (n (n + 1))^{2 p}},
& \quad m < M_\theta / 2, \\
0, & \quad \text{otherwise},
\end{cases}
\end{equation}
where $\alpha \ll 1$ is a real constant and $p \in \mathbb{N}$ is an integer
power. We can see here that the orders $m$ are filtered using an ideal
filter and the degrees $n$ are weighted in terms of the parameters $\alpha$ and
$p$. This weighting is essentially a form of Tikhonov regularization
(see~\cite{Mcclarren2010}) which minimizes
\[
\frac{1}{2} \int_{\mathbb{S}^2} |f - F|^2 \dx[S]
+ \frac{\alpha}{2} \int_{\mathbb{S}^2} |\Delta^p_{\mathbb{S}^2} F|^2 \dx[S],
\]
where $f$ is the function in question an $F$ is its projection to the
spherical harmonic basis. The desired expression can then be obtained by
using the fact that the spherical harmonics are eigenfunctions of the
Laplace-Beltrami operator on the sphere, i.e.
\[
\Delta_{\mathbb{S}^2} Y^m_n = -n (n + 1) Y^m_n.
\]

In practice, we have found that $p = 2$ and $\alpha \approx 10^{-5}$ provides
sufficient filtering to stabilize the droplet evolution. These parameters can
also be determined directly from the data by, e.g., minimizing a Generalized
Cross-Validation functional~\cite{Golub1979}. The filtered coefficients
of the normal of a unit sphere can also be seen in~\Cref{fig:normal}.

\subsection{Time Stepping and Adaptation}
\label{ssc:methods:timestepping}

We discretize the evolution equation~\eqref{eq:evolution} with the
kinematic condition~\eqref{eq:kinematic} using a second-order explicit method
\begin{equation} \label{eq:ssp_rk2}
\begin{aligned}
\tilde{\vect{X}}^{n + 1} =\,\, &
    \vect{X}^n + \Delta t^n \vect{V}(t^n, \vect{X}^n), \\
\vect{X}^{n + 1} =\,\, & \frac{1}{2} \vect{X}^n
+ \frac{1}{2} \left\{
\tilde{\vect{X}}^{n + 1} + \Delta t^n \vect{V}(t^{n + 1}, \tilde{\vect{X}}^{n + 1})
    \right\},
\end{aligned}
\end{equation}
where both the velocity field and the geometry in the source term are computed
at the time $t^n$. We have found that higher-order time integration does not
offer significant improvement in stability and volume conservation. However,
the cost of multiple solutions to~\eqref{eq:density} can quickly dominate.
This is more important in optimization, where we must solve the geometry evolution
equation and the normal adjoint transverse field equation a potentially
large number of times. For comparison, in~\cite{Ojala2015} (2D), the authors
have used an embedded second-order method and in~\cite{Veerapaneni2011} (3D)
first-order explicit and implicit schemes have been analyzed. In~\cite{Fikl2021},
we have analyzed an axisymmetric formulation coupled with a third-order SSP RK3
time integrator to match the spatial discretization order. The time step is
fixed, but proportional to a characteristic mesh size
\[
\Delta t = C h_{max},
\quad \text{where }
h_{\max} = \max_{i \in \{0, \dots, M\}} \max(\sigma_{i, 1}, \sigma_{i, 2}),
\]
where $\sigma_i$ are the eigenvalues of the metric tensor at each point
$\vect{X}_i(0)$. This approximation for a characteristic mesh length is
suited for scenarios where the elements can be highly skewed, e.g. at the poles.
Improved time step adaptation for long-time simulations can
be found, e.g., in~\cite{Zinchenko2013}.

The largest issue when performing the droplet evolution, for medium to long
time spans, is the potentially rapid degradation of the mesh. Even for a
single droplet in a uniform flow, using the kinematic condition~\eqref{eq:kinematic}
without $\vect{w}$ will cluster the points in the downstream region of the droplet.
Therefore, it is necessary to implement a method that either maintains or improves
the mesh quality during the evolution. In our implementation, we prefer to make
use of so-called ``passive'' mesh adaptation methods, where a tangential
velocity field is added as in~\eqref{eq:kinematic}

This tangential field is chosen in such a way that, over several iterations,
it can improve the quality of the mesh. For two-dimensional flows, the methods
presented in~\cite{Kropinski2001} are a robust solution that preserves arclength
or clusters points in regions of higher curvature. However, in three-dimensional
flows, it becomes increasingly difficult to develop robust methods that maintain
the mesh quality. We use the tangential velocity field described
in~\cite{Zabarankin2013}. In this method, we find a velocity field that
(approximately) minimizes
\begin{equation} \label{eq:tangential}
F(\vect{w}) = \frac{\alpha}{2} \sum_{ij} \left[
\od{}{t} \left(\frac{\|\vect{x}_{ij}\|^2}{h_{ij}^2}
+ \frac{h_{ij}^2}{\|\vect{x}_{ij}\|^2}\right)
\right]^2
+ \frac{\beta}{2} \sum_{m} \frac{1}{C_m^2} \left[
\od{C_m}{t}
\right]^2,
\end{equation}
where $\alpha, \beta \in \mathbb{R}_+$. Here, $\vect{x}_{ij}$ denotes the
edge between the vertices $(\vect{x}_i, \vect{x}_j)$, $h_{ij}$ is a
corresponding curvature-based length scale and $C_m$ measures the ratio between
the element area and its edge lengths. In~\ref{ax:adaptive}, we define the
terms in this cost functional and give appropriate extensions to quadrilateral
elements (as~\cite{Zinchenko2013} is focused on simplices). As we will not
be stress testing this method with high deformations, we have found it
sufficient to ensure the mesh maintains sufficient quality and the simulation
remains stable for longer times. A summary of all the steps required to solve
the forward problem are given in~\Cref{alg:stokes}.

\begin{figure}[ht!]
\centering
\begin{minipage}{0.8\linewidth}
\begin{algorithm}[H]
\caption{Two-phase Stokes Interface Evolution.}
\label{alg:stokes}
\KwData{Number of elements $(M_\theta, M_\phi)$;
    Spherical harmonic expansion orders $(N_\theta, N_\phi)$;
    Gauss-Legendre quadrature order $P$ and $P_{quad}$;
    QBX order $P_{qbx}$; FMM order $P_{fmm}$.}
\KwData{Final time $T$; time step $\Delta t$ or Courant number $C$.}
\KwData{Initial parametrization $\vect{X}_0$ for the surface $\Sigma(0)$.}
\ForEach{$n \in \{0, \dots, N - 1\}$}{
\begin{enumerate}[1.]
    \item Solve for the density $\vect{q}^n$ in~\eqref{eq:density} using the
    boundary conditions of~\eqref{eq:stokes}.
    \item Compute the velocity field $\vect{u}^n$ using~\eqref{eq:velocity}.
    \item Compute the tangential velocity field $\vect{w}^n$ solving~\eqref{eq:tangential}.
    \item[*.] Checkpoint solutions $(\vect{q}^n, \vect{X}^n)$ for the adjoint problem.
    \item Evolve interface in time using~\eqref{eq:ssp_rk2} and apply the
    filter~\eqref{eq:filter} to the right-hand side at each stage.
\end{enumerate}
}

Set cost as $J = y^{N}$.
\end{algorithm}
\end{minipage}
\end{figure}

As part of the optimization problem, we must also compute the cost~\eqref{eq:cost}.
This is generally done by noting that
\[
J = \int_0^T j(\vect{u}, \vect{X}, \vect{g}) \dx[t],
\]
can also be understood as an ODE. In particular, let
\begin{equation} \label{eq:cost_ode}
\begin{cases}
\dot{y}(t) = j(\vect{u}, \vect{X}, \vect{g}), \\
y(0) = 0,
\end{cases}
\end{equation}
then we have that $J = y(T)$. In the form of an ODE, we can make use of the
same time integration method as for the equation themselves and obtain an accurate
approximation of time-dependent cost functionals.

\subsection{Adjoint Discretization}
\label{ssc:methods:adjoint}

In the previous sections, the focus has been exclusively on the state equations
and the evolution of the droplets. For the adjoint system, we have a similar
set of equations to solve. First, we can see from~\eqref{eq:adjoint:stokes}
and~\eqref{eq:adjoint:jumps} that the adjoint Stokes equations have the exact
same structure as the state equations. Therefore, the same solver can be used
in both cases, while only accounting for the differences in boundary conditions.
Then, we are left with discretizing the normal adjoint transverse equation
from~\eqref{eq:normal_adjoint_transverse_field}. In this case, we simply
write
\[
-\od{X^*}{t} = V^*[t, X^*; \vect{X}],
\]
which is discretized with the dual consistent scheme
\begin{equation} \label{eq:ssp_rk2_adjoint}
\begin{aligned}
\tilde{X}^{*, n} = \,\, &
X^{*, n + 1} + \Delta t^{n + 1} V^*[t^{n + 1}, X^{*, n + 1}; \tilde{\vect{X}}^{n}], \\
X^{*, n} =\,\, & \frac{1}{2} X^{*, n + 1} + \frac{1}{2} \left\{
\tilde{X}^{*, n} + \Delta t^{n + 1}
V^*[t^n, \tilde{X}^{*, n}; \vect{X}^n]
\right\},
\end{aligned}
\end{equation}
where the state and adjoint variables are not computed at the same times.
We express the terms in right-hand side $V^*$ in a strong form through the
layer potentials defined in~\Cref{ssc:methods:qbx} and all their derivatives. A
complete description of the steps is given in~\Cref{alg:adjoint}. As in the
case of~\Cref{alg:stokes}, an approximation of the gradient can be incorporated
if the control is not time-dependent.

\begin{figure}[ht!]
\centering
\begin{minipage}{0.8\linewidth}
\begin{algorithm}[H]
\caption{Normal Adjoint Transverse Field Evolution.}
\label{alg:adjoint}
\KwData{Solutions $(\vect{q}^n, \vect{X}^n)$ of the forward problem for all $n$.}
\ForEach{$n \in \{N - 1, \dots, 1\}$}{
\begin{enumerate}[1.]
    \item Solve for the density $\vect{q}^{*, n + 1}$ in~\eqref{eq:density} using
    the boundary conditions of the adjoint problem~\eqref{eq:adjoint:stokes}.
    \item Compute the adjoint velocity field $\vect{u}^{*, n + 1}$
    using~\eqref{eq:velocity}.
    \item Evolve the normal adjoint transverse field using~\eqref{eq:ssp_rk2_adjoint}
    by applying the filter~\eqref{eq:filter} to the right-hand side at each stage.
    \item[.*] If the control is time dependent, then use~\eqref{eq:cg_update}
    to update the components at time $t^n$. Otherwise, evolve an ODE
    similar to~\eqref{eq:cost_ode} to accumulate the final gradient.
\end{enumerate}
}
\end{algorithm}
\end{minipage}
\end{figure}

\begin{remark}
When using a discretize-then-optimize approach to adjoint optimization, the
forward and adjoint problems have exactly the same time step requirements
(as the operators in question are just Hermitian transposes of each other).
However, for an optimize-then-discretize approach this is no longer the case
and additional errors can appear due to a lack of dual consistency. We have
not found this to be an issue in our simulations, so the same time step is
used.
\end{remark}

\subsection{Gradient Descent}
\label{ssc:methods:descent}

The optimization problem~\eqref{eq:optim} can now be solved by standard
gradient descent methods. In particular, we can use~\Cref{alg:stokes}
and~\Cref{alg:adjoint} as black boxes used to compute the cost functional
values and the corresponding gradient at every $\vect{g}^{(k)}$. Then,
a steepest descent update is applied
\begin{equation} \label{eq:cg_update}
\vect{g}^{(k + 1)} = \vect{g}^{(k)} + \alpha^{(k)} \nabla_{\vect{g}} J(\vect{g}^{(k)}),
\end{equation}
where $\alpha^{(k)}$ is a chosen step size. In this work, we use the
Barzilai-Borwein method from~\cite{Burdakov2019} to approximate a value for
$\alpha^{(k)}$ at every step of the optimization. If this value does not result
in a decrease of the cost functional, we set $\alpha^{(k)} \gets \alpha^{(k)} / 2$
until a decrease is achieved. This approximation was chosen to avoid a
more complicated line search, as evaluating the cost functional through
\Cref{alg:stokes} can be very costly. The complete method is described
in~\Cref{alg:descent}.

\begin{figure}[ht!]
\centering
\begin{minipage}{0.9\linewidth}
\begin{algorithm}[H]
\caption{Adjoint-based Optimization of Quasi-static Two-Phase Stokes Flow.}
\label{alg:descent}
\KwData{Initial guess $\vect{g}^{(0)}$ for the control.}
\KwData{Tolerances for the cost functional values $\epsilon_J$, the gradient
    $\epsilon_\nabla$ and the step size $\alpha_{max}$.}

\ForEach{$k \in {0, \dots, K}$}{
\begin{enumerate}[1.]
    \item Compute the cost functional $J(\vect{g}^{(k)})$ using~\Cref{alg:stokes}.
    \item Compute the gradient $\nabla_{\vect{g}} J(\vect{g}^{(k)})$
    using~\Cref{alg:adjoint}.
    \item Estimate $\alpha^{(k)}$ using the Barzilai-Borwein method
    bounded by $\alpha_{max}$.
    \item Update the control
    \[
    \vect{g}^{(k + 1)} = \vect{g}^{(k)} + \alpha^{(k)} \nabla_{\vect{g}} J(\vect{g}^{(k)}),
    \]
    \item[*.] If $J(\vect{g}^{(k + 1)}) > J(\vect{g}^{(k)})$, attempt the update
    again with $\alpha^{(k)} / 2$.
    \item Check stopping criteria
    \[
    |J(\vect{g}^{(k + 1)})| < \epsilon_J
    \quad \text{or} \quad
    \|\nabla_{\vect{g}} J(\vect{g}^{(k + 1)})\| < \epsilon_\nabla.
    \]
\end{enumerate}
}
\end{algorithm}
\end{minipage}
\end{figure}

\section{Numerical Results}
\label{sc:results}

We present several static and quasi-static test cases to serve as verification
for the adjoint-based optimal control of two-phase Stokes flows presented in
the previous sections. For simplicity, we fix a set of discretization
parameters in~\Cref{tbl:parameters}. The number of elements and spherical
harmonic orders will be case-dependent.

\begin{table}[ht!]
\centering
\begin{tabular}{ll|c} \toprule
Parameter & Description & Value \\\midrule
$P$ & Base quadrature rule order & 3 \\
$P_{quad}$ & Oversampled quadrature rule order & $4 P$ \\
$P_{qbx}$ & QBX expansion order & 4 \\
$P_{fmm}$ & FMM expansion order & 10 \\
\bottomrule
\end{tabular}
\caption{Default discretization parameters used in Test 1-3}
\label{tbl:parameters}
\end{table}

\subsection{Test 1: Verification of the Adjoint Gradient}
\label{ssc:results:1}

As a first test, we will present some results to verify the adjoint equations
presented in~\Cref{sc:control}. The standard method to verify a
continuous adjoint in the optimize-then-discretize setting is by comparing
the gradient to a black-box finite different approximation. For our work, it
is very difficult to obtain clear convergence results, due in part to the
moving geometry and the complexity of the source terms in the adjoint
transverse field equation~\eqref{eq:transverse_field_equation}. Therefore, in
this test we remove the moving geometry and focus on verifying the right-hand
side source terms. As shown in~\cite{Fikl2021}, these terms are essentially
part of the shape gradient of the static problem. In our cost
functional~\eqref{eq:cost}, we take
\[
J \triangleq \frac{1}{2} \int_{\Sigma} (\vect{u} \cdot \vect{n})^2 \dx[S],
\]
i.e. $\alpha_1 = 1, \alpha_2 = \alpha_3 = 0$ and $u_d = 0$.
To compute a finite difference approximation of the shape gradient, we use
\begin{equation} \label{eq:finite}
g_{FD, i} \triangleq \inp{\vect{g}, \vect{h}_i}_\Sigma \approx
\frac{J(\vect{X} + \epsilon \vect{h}_i) - J(\vect{X})}{\epsilon},
\end{equation}
where $\vect{g}$ is the gradient and $\vect{h}_i \triangleq h_i \vect{n}$
is a chosen bump function. Typically in finite different approximations, $h_i$
is chosen to be a Dirac Delta point mass, but this is not possible here as
we require a smooth perturbation of the geometry $\vect{X}$. We use
\[
h_i(\vect{X}) = \frac{1}{\sqrt{2 \pi} \sigma} \exp \left(
-\frac{\|\vect{X} - \vect{X}_i\|}{2 \sigma^2}
\right),
\]
where the standard deviation $\sigma$ is taken to be $10^{-1}$. In general, a
good choice of $\sigma$ depends on the mesh spacing. As the bump functions are
not given pointwise, to compare the finite different approximation with the
adjoint-based approximation, we must also convolve the adjoint based gradient
\[
g_{AD, i} \triangleq \inp{\nabla_{\vect{X}} J, \vect{h}_i}_\Sigma,
\]
where the inner product is computed using the quadrature rules of the Nyström-QBX
method from~\Cref{ssc:methods:qbx}. The errors are given by
\[
E[\vect{g}_{AD}, \vect{g}_{FD}] \triangleq
\frac{\|\vect{g}_{AD} - \vect{g}_{FD}\|_2}{\|\vect{g}_{FD}\|_2},
\]
in the standard square-summable $\ell^2$ norm. It would be prohibitively
expensive to compute a perturbation for the finite difference approximation at
all the points in the geometry, especially on finer meshes. We have chosen
here $5$ random points on a sphere of radius $R = 1$ at which to compute
the finite difference gradient and the errors.

\begin{table}[ht]
\centering
\begin{tabular}{l|ll|ll} \toprule
$\epsilon$ & $(\lambda = 1, \mathrm{Ca} = 1)$ & \textbf{EOC}
& $(\lambda = 10, \mathrm{Ca} = \infty)$ & \textbf{EOC} \\ \midrule
1.00e-01 & 3.886568e-02 & ---  & 1.221459e-01 & ---  \\
1.00e-02 & 4.032369e-03 & 0.98 & 1.348132e-02 & 0.96 \\
1.00e-03 & 6.707820e-04 & 0.78 & 1.747667e-03 & 0.89 \\
1.00e-04 & 5.016154e-04 & 0.13 & 6.727348e-04 & 0.41 \\
1.00e-05 & 4.956386e-04 & 0.01 & 5.883712e-04 & 0.06 \\
\bottomrule
\end{tabular}

\caption{
    Test 1: Relative errors for the finite difference vs adjoint gradient
    approximations for $(\lambda, \mathrm{Ca})$ as $(1, 1)$ and $(10, \infty)$.}
\label{tbl:test1:convergence}
\end{table}

\begin{figure}[ht!]
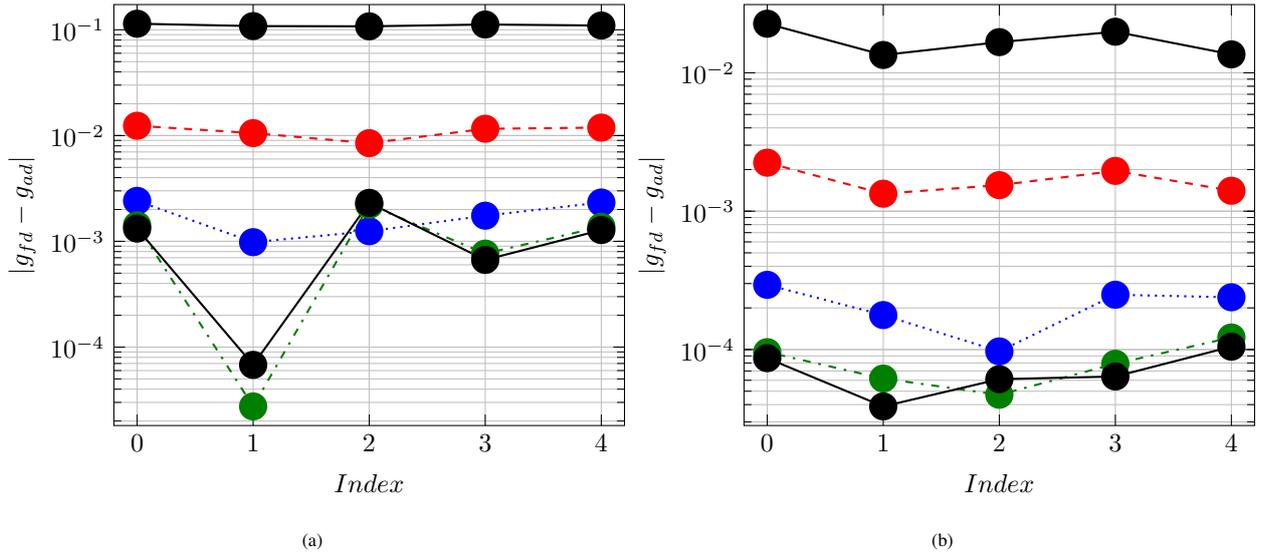

\begin{subfigure}[b]{0.5\linewidth}
\centering
\includefigure{data_convergence_capillary_errors}
\caption{}
\end{subfigure}
\hfill
\begin{subfigure}[b]{0.5\linewidth}
\centering
\includefigure{data_convergence_viscosity_errors}
\caption{}
\end{subfigure}
\caption{
    Test 1: Absolute errors for the finite difference vs adjoint gradient
    approximations for $(\lambda, \mathrm{Ca})$ as $(1, 1)$ and $(10, \infty)$.}
\label{fig:test1:convergence}
\end{figure}

For the discretization, we have chosen a sufficiently fine grid to showcase
the first-order convergence of the finite difference approximation~\eqref{eq:finite}.
In general, we expect that the error has two parts $\mathcal{O}(\epsilon) +
\mathcal{O}(h_{max}^P)$. For this experiment, as we vary $\epsilon$, we expect
to see first-order convergence until a plateau determined by the discretization
error that depends on $h_{max}$. In~\cite{Fikl2021}, we have shown convergence
with respect to the mesh spacing as well. The discretization in this case
is formed out of $M_{\theta} = M_{\phi} = 64$ elements and
$N_\theta = N_{\phi} = 94$ spherical harmonic coefficients. No additional
filtering is performed on the gradient. For the two-phase Stokes problem, we
have used $\lambda = 1$ and $\mathrm{Ca} = 1$ as base parameters. Then, the
farfield boundary conditions are given by
\begin{equation} \label{eq:extensional}
\vect{u}_\infty = (\alpha x, -2 \alpha y, \alpha z)
\quad \text{and} \quad
p_\infty \equiv 0,
\end{equation}
i.e. an extensional flow along the $y$ axis. This choice was made to match
the axisymmetric extensional flow used in~\cite{Fikl2021}. The resulting
relative errors can be seen in~\Cref{tbl:test1:convergence}. The first-order
convergence is recovered for both $\mathcal{S}^*$ and $\mathcal{C}^*$
terms in the adjoint-based gradient (see~\eqref{eq:adjoint_transverse_field}).
For completeness, the pointwise errors can be seen in~\Cref{fig:test1:convergence}
at the chosen points
\[
(m, p) \in \{
(3652, 2), (3574, 9), (2723, 7), (1614, 3), (322, 1)
\},
\]
where $m$ denotes the element and $p$ denotes the local node index inside the
element.

\subsection{Test 2: Steady State Tracking}
\label{ssc:results:2}

Next, we will look at a simple shape optimization problem using the gradient
that we have verified in the previous section. The problem involves finding
a steady state of the Stokes flow in a particular configuration. The setup
we will be following is presented in~\cite{Stone1989}, where a single droplet
has been suspended in an axisymmetric extensional flow. We will attempt to find
the steady state for $(\lambda = 1, \mathrm{Ca} = 0.05)$ in the extensional
flow~\eqref{eq:extensional} and match the results from~\cite{Stone1989}.
The cost functional we will be using is
\begin{equation} \label{eq:steady_cost}
J = \frac{1}{2} \int_{\Sigma} (\vect{u} \cdot \vect{n})^2 \dx[S]
+ \frac{1}{2} (V - V_d),
\end{equation}
where the volumes
\[
V_d \triangleq \frac{4 \pi}{3} R^3
\quad \text{and} \quad
V \triangleq \int_{\Omega_-} \dx[S]
    = \frac{1}{d} \int_{\Sigma} \vect{x} \cdot \vect{n} \dx[S]
\]
are used to enforce volume conservation in the optimization problem.
Note that, unlike the kinematic condition~\ref{eq:kinematic}, the shape
optimization problem will not attempt to keep a constant volume. The
discretization from~\cref{ssc:results:1} is used.

With this setup in mind, we perform two separate optimization problems to
gauge the benefit of using the full power of shape optimization on the
cost functional from~\eqref{eq:steady_cost}. First, we start by
finding an optimal spheroid that matches our volume and minimizes
the normal velocity $\vect{u} \cdot \vect{n}$. This problem has been studied
in~\cite{Zabarankin2013}, where an analytical solution is found on a prolate
spheroid. This analytical solution is then used as part of an optimization
problem to determine the parametrization of the spheroid that minimizes
$\vect{u} \cdot \vect{n}$, in the same way that we are attempting to do here.
In~\cite{Zabarankin2013}, the authors perform this optimization for a wide range
of $(\lambda, \mathrm{Ca})$ and compare the results to existing literature. We
will restrict ourselves to the single case with $\lambda = 1$ and $\mathrm{Ca} = 0.05$.
Before we can continue, we detail the parametrization of the spheroid and how
the adjoint-based gradient can be computed with respect to the new parameters.
First, we have that the spheroid is given by
\[
\vect{S}(\theta, \phi; R, a, \alpha, \beta, \gamma) \triangleq
\vect{R}(\alpha, \beta, \gamma)
\begin{bmatrix}
R \sin \theta \cos \phi \\
R \sin \theta \sin \phi \\
a R \cos \theta,
\end{bmatrix}
\]
where $R$ is the radius and $a$ is the aspect ratio of the spheroid. The
rotation matrix is given in terms of the Euler angles as
\[
\vect{R}(\alpha, \beta, \gamma) = \vect{R}_z(\alpha) \vect{R}_y(\beta) \vect{R}_x(\gamma),
\]
where each $\vect{R}_i$ matrix rotates around the $i$ axis. Then, we use the
parameters $\vect{s} \triangleq (R, a, \alpha, \beta, \gamma) \in
\mathbb{R}_+^2 \times \mathbb{R}^3$ as our control variables and strive to find the
gradient of the cost functional from~\eqref{eq:steady_cost} with respect to
$\vect{s}$ instead of $\vect{X}$. The gradient is obtained by a straightforward
application of the chain rule as
\[
\pd{J}{s_i} = \int_\Sigma
    \nabla_{\vect{X}} J \vect{n} \cdot \pd{\vect{S}}{s_i} \dx[S],
\]
where
\[
\pd{\vect{S}}{R} = \frac{1}{R} \vect{S},
\qquad
\pd{\vect{S}}{a} = \vect{R}(\alpha, \beta, \gamma)
\begin{bmatrix}
0 & 0 & 0 \\
0 & 0 & 0 \\
0 & 0 & 1/a
\end{bmatrix}
\vect{R}^T(\alpha, \beta, \gamma) \mathbf{S},
\]
and the derivative with respect to the Euler angle $\alpha$ is given by
(similarly for $\beta, \gamma$)
\[
\pd{\vect{S}}{\alpha} = \pd{\vect{R}}{\alpha} \vect{R}^T \vect{S},
\quad \text{where } \quad
\pd{\vect{R}}{\alpha} = \od{\vect{R}_z}{\alpha} \vect{R}_y(\beta) \vect{R}_x(\gamma)
=
\begin{bmatrix}
-\sin \alpha & -\cos \alpha & 0 \\
\cos \alpha & -\cos \alpha & 0 \\
0 & 0 & 0
\end{bmatrix}
\vect{R}_y(\beta) \vect{R}_x(\gamma).
\]

\begin{remark}
The gradient described above is not the complete gradient with respect to
$\vect{s}$ that we expect. The discrepancy comes from the fact that, in
obtaining the shape gradient, we have assumed that the perturbations
$\tilde{\vect{V}}$ only have a normal component (see~\eqref{eq:shape_derivatives}).
However, for this problem, the perturbations are actually given by
\[
\tilde{\vect{V}} \triangleq \od{\vect{S}}{s_i},
\]
which have tangential components.
\end{remark}

\begin{figure}[ht!]
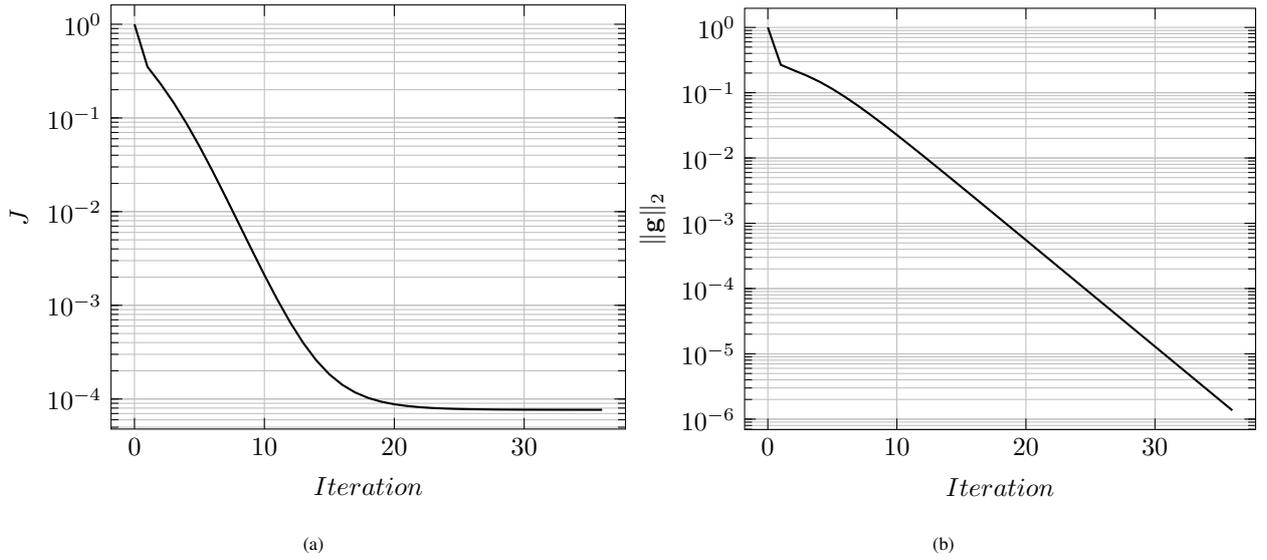

\begin{subfigure}[b]{0.5\linewidth}
\centering
\includefigure{optim_qbx_shape_spheroid_cost}
\caption{}
\end{subfigure}
\hfill
\begin{subfigure}[b]{0.5\linewidth}
\centering
\includefigure{optim_qbx_shape_spheroid_gradient}
\caption{}
\end{subfigure}
\caption{Test 2: (a) Cost functional and (b) Gradient $\ell^2$ norm for the
    spheroid optimization, normalized with respect to initial guess.}
\label{fig:test2:sphereoid}
\end{figure}

For the optimization parameters, we start with a sphere of radius $R = 1$, much
like we would start the genuine evolution equation~\eqref{eq:evolution}
in an attempt to find the steady state. Therefore, for the spheroid parameters,
we take $\vect{s}^{(0)} \triangleq (1, 1, 0, 0, 0)$, while the full shape
optimization will use $\vect{X}^{(0)} \triangleq \vect{S}(\theta, \phi;
\vect{s}^{(0)})$. We have performed optimizations with non-zero angles
$(\alpha, \beta, \gamma)$ and observed similar results, but for the sake of
comparison, we reduce here to this simple case.

\begin{figure}[ht!]
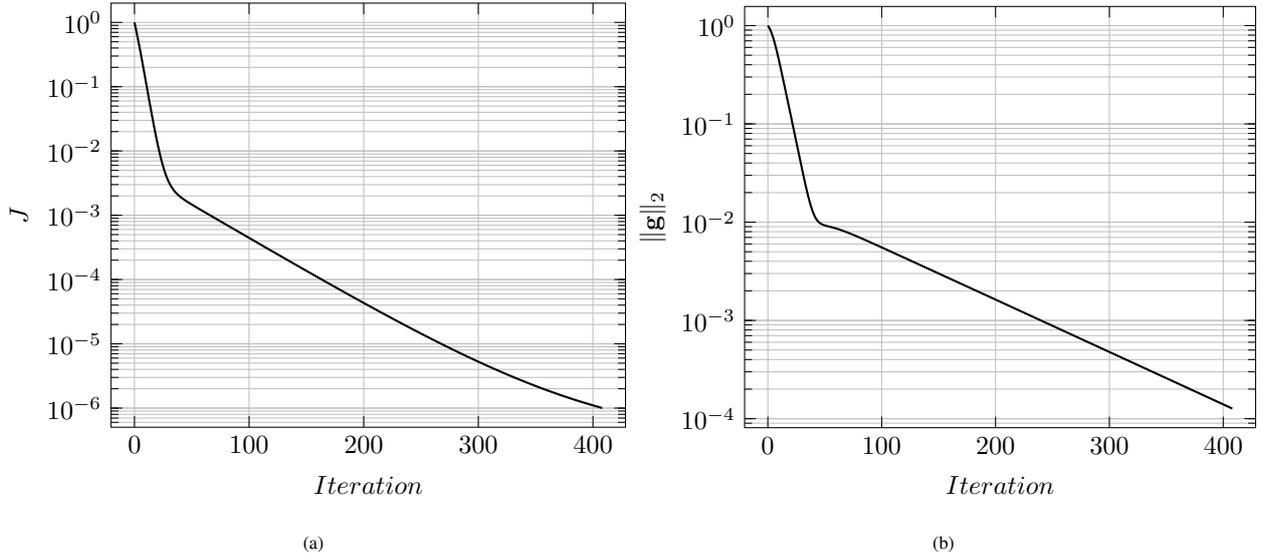

\begin{subfigure}[b]{0.5\linewidth}
\centering
\includefigure{optim_qbx_shape_3d_cost}
\caption{}
\end{subfigure}
\hfill
\begin{subfigure}[b]{0.5\linewidth}
\centering
\includefigure{optim_qbx_shape_3d_gradient}
\caption{}
\end{subfigure}
\caption{Test 2: (a) Cost functional and (b) Gradient $\ell^2$ norm for the
    shape optimization, normalized with respect to initial guess.}
\label{fig:test2:convergence}
\end{figure}

The results for the spheroid optimization are provided
in~\Cref{fig:test2:sphereoid}. We can see that the cost functional
plateaus at $J \approx 10^{-4}$, while the gradient is about two orders
of magnitude smaller. This indicates that we have indeed found a minimum for
this optimization problem, even if it is not $J \approx 0$. A clear benefit
of this choice is that the optimization only required $36$ steps to achieve
convergence and find a shape with $\|\vect{u} \cdot \vect{n}\|_2 \approx 10^{-2}$.
In many situations of interest this is a sufficiently accurate result.

In the spheroid optimization, the cost functional cannot reach the minimum
solution as the steady state of a droplet in extensional flow is not exactly
a spheroid. Therefore, as a second experiment, we attempt to perform a full
shape optimization and see if better convergence can be obtained. This is
indeed the case, as shown in~\Cref{fig:test2:convergence}. Here, the cost
functional reached a minimum value of $10^{-6}$, two orders of magnitude smaller
than in the spheroid case. Therefore, if we want to find a true steady state
shape to arbitrary precision, the full shape optimization problem is required.
The downside, however, is that the optimization required roughly $400$ steps
to reach the minimum, making it significantly more costly, i.e. $13 \times$ for an
increase in accuracy of two orders of magnitude. This is more in line with
the requirements of computing the steady state using~\eqref{eq:kinematic}.

\begin{figure}[ht!]
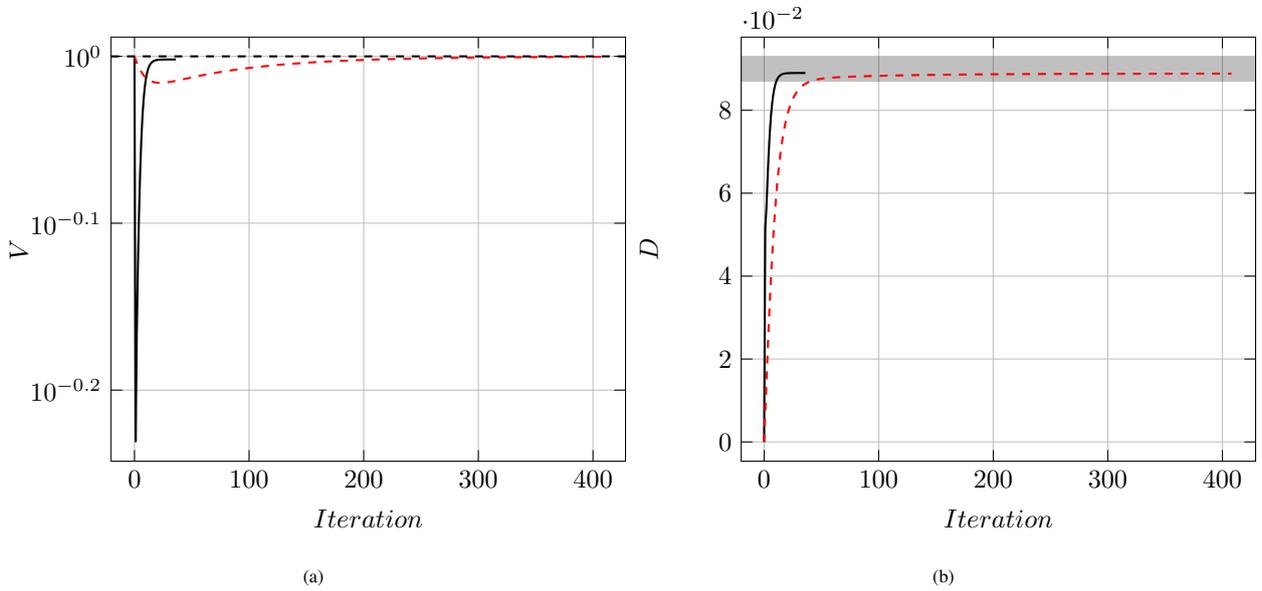

\begin{subfigure}[b]{0.5\linewidth}
\centering
\includefigure{optim_qbx_shape_spheroid_volume}
\caption{}
\end{subfigure}
\hfill
\begin{subfigure}[b]{0.5\linewidth}
\centering
\includefigure{optim_qbx_shape_spheroid_deformation}
\caption{}
\end{subfigure}
\caption{Test 2: (a) Droplet volume $|\Omega_-|$ and (b) deformation during
the optimization for the spheroid (full) and shape (dashed) optimization.}
\label{fig:test2:deformation}
\end{figure}

To see if the optimization reached the expected steady state shape, we briefly
investigate the volume and deformation during the optimization for both cases.
First, we can see the improved behavior in the convergence of the volume and
deformation of the droplet in~\Cref{fig:test2:deformation}.
Following~\cite{Stone1989}, the deformation is defined simply as
\[
D \triangleq \frac{L - H}{L + H},
\]
where $L$ is the half-length and $H$ is the half-height of the droplet in
the $x$-$z$ plane. An approximation of the results obtained in~\cite{Stone1989}
is also provided (the exact values are not available). While it is hard to
determine an improvement in the deformation, we can clearly see that the
volume is not accurately conserved when performing the spheroid optimization,
as there is a noticeable gap in~\Cref{fig:test2:deformation}a. This is
due to the fact that the velocity term in the cost functional is dominating
and did not allow convergence of the smaller volume term.

\subsection{Test 3: Unsteady Helix}
\label{ssc:results:3}

\begin{figure}[ht!]
\centering
\begin{tikzpicture}
\begin{scope}[rotate=-90]
\newcommand{\modhelix}[3][4]{% #1=number of cycles, #2=modulus, #3=start value
\foreach \k in {1,...,#1}{%
    \foreach \j [evaluate=\j as \m using (\j-1)/#2*360+90/#2,
                 evaluate=\j as \n using \j/#2*360+90/#2,
                 evaluate=\j as \c using int(#3+(#2*(\k-1)+\j-1))] in {1,...,#2}{%
    \draw[dashed, white, double=black, double distance=1.5pt, opacity=0.25]
        ({2*cos(\m)}, {2*\k + sin(\m) + \m/180 - 1.3}) -- ++ (0, 2);
    \draw[white, double=JCPLightBlue, double distance=1.5pt, line width=1pt, domain=\m-90/#2:\n-90/#2, smooth, variable=\t]
        plot ({2*cos(\t)}, {2*\k+sin(\t)+\t/180});
    \fill ({2*cos(\m)}, {2*\k + sin(\m) + \m/180}) circle [radius=3pt];
}}}

\modhelix[4]{5}{-8};
\end{scope}

\draw[-latex] (-1, 0) -- (0, 0) node [below] {$z$};
\draw[-latex] (-1, 0) -- (-1, 1) node [right] {$y$};
\draw[-latex] (-1, 0) -- (-1.6, -0.6) node [right] {$x$};
\end{tikzpicture}

\caption{Test 3: Helical trajectory of a particle.}
\label{fig:test3:geometry}
\end{figure}
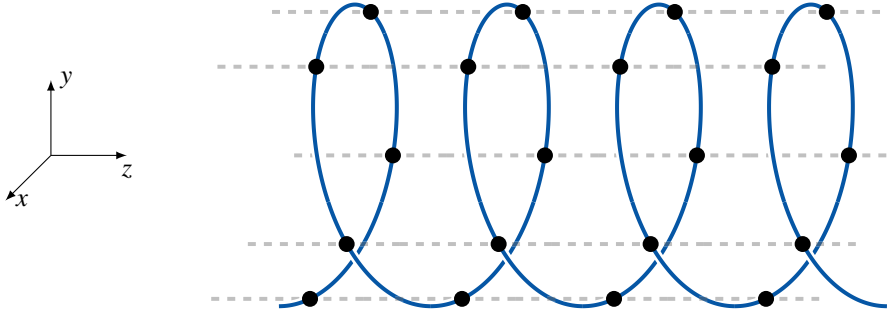

For a quasi-static optimization problem, we attempt to match the
movement of two droplets against a desired helical trajectory. The cost
functional we will be using is given by
\[
J = \frac{1}{2} \sum_{i \in \{+, -\}}
\int_0^T \int_{\Sigma_i} \|\vect{x}_{c, i}(t) - \vect{x}_{c, d, i}(t)\|^2 \dx[S] \dx[t],
\]
where $\vect{x}_{c, \pm}$ are the centroids of the two droplets. The desired
centroids are taken as the centroids of two spheres of radius $R$, i.e.
\begin{equation} \label{eq:test3:xd}
\vect{x}_{c, d, \pm}(t; \omega_d, H_d) =
\begin{bmatrix}
\cos \omega_d t / T & -\sin \omega_d t / T & 0 \\
\sin \omega_d t / T & \cos \omega_d t / T & 0 \\
0 & 0 & 1
\end{bmatrix}
\begin{bmatrix}
    \pm 2 R \\
    0 \\
    0
\end{bmatrix}
+
\begin{bmatrix}
    0 \\
    0 \\
    H_d t / T
\end{bmatrix},
\end{equation}
in the time interval $t \in [0, T]$, where $\omega_d, H_d \in \mathbb{R}_+$
denote the wavenumber and the height of the helix, respectively. The initial
conditions $\vect{X}_{0, \pm}(\theta, \phi)$ are also taken to be spheres of
radius $R = 1$. The velocity field that gives rise to the trajectory
from~\eqref{eq:test3:xd} is essentially a solid body rotation in the $x$-$y$
plane and a uniform flow along the $z$ axis. We will use a parametrization
in terms of $(\omega, H)$ of the farfield as our control variables. As such,
the farfield boundary conditions are given by
\[
\vect{u}_\infty(t, \vect{x}; \omega, H) \triangleq
-\frac{\omega}{T} y \vect{e}_x
+ \frac{\omega}{T} x \vect{e}_y + \frac{H}{T} \vect{e}_z
\quad \text{and} \quad
p_\infty(t, \vect{x}; \omega, H) \equiv 0,
\]
which will result in a classical helical trajectory seen
in~\Cref{fig:test3:geometry}. To complete the definition of the two-phase
Stokes problem, we again choose $\lambda = 1$ and $\mathrm{Ca} = 0.05$. In this
case, the adjoint-based gradient is given by
\[
\begin{aligned}
\pd{J}{\omega} = \,\, & \frac{1}{T} \int_0^T
    \int_{\Sigma(t)} (\vect{X}^* \cdot \vect{n}) (x n_y - y n_x) \dx[S]
    \dx[t], \\
\pd{J}{H} = \,\, & \frac{1}{T} \int_0^T
    \int_{\Sigma(t)} (\vect{X}^* \cdot \vect{n}) n_z \dx[S]
    \dx[t].
\end{aligned}
\]

\begin{figure}[ht!]
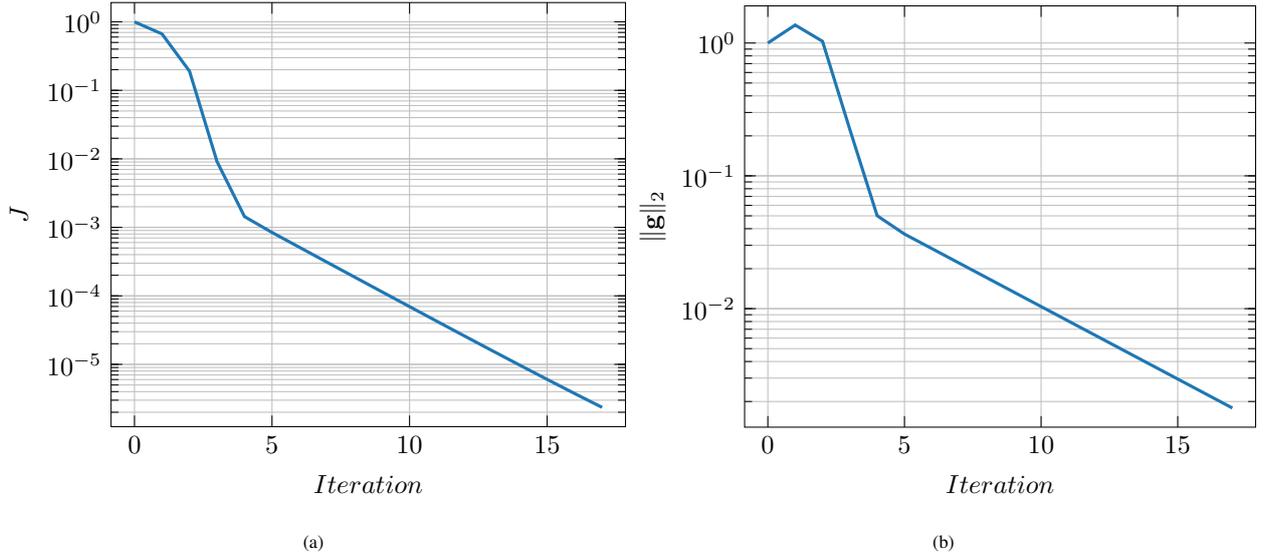

\begin{subfigure}[b]{0.5\linewidth}
\centering
\includefigure{optim_qbx_helix_cost}
\caption{}
\end{subfigure}
\hfill
\begin{subfigure}[b]{0.5\linewidth}
\centering
\includefigure{optim_qbx_helix_gradient}
\caption{}
\end{subfigure}
\caption{Test 3: Normalized (a) cost functional values and (b) gradient norm
values.}
\label{fig:test3:convergence}
\end{figure}

The spatial discretization of the droplets is similar to that of the previous
example. We set $M_{\theta} = M_{\phi} = 24$ elements and $N_\theta = N_\phi =
45$ spherical harmonic coefficients. As mentioned in~\Cref{ssc:methods:timestepping}
and~\Cref{ssc:methods:adjoint}, we apply a filter at every step of the evolution,
both for the forward and the adjoint calculations. The droplets are evolved
using the second-order SSP RK time integrator to a final time of $T = \pi$ with a
fixed time step of $\Delta t = 10^{-2}$. The time step was chosen to ensure
stability during the evolution of the geometry using the desired control
$(\omega_d, H_d)$. We make use of the tangential forcing
from~\Cref{ssc:methods:timestepping} to ensure the mesh does not significantly
degrade during the simulation.

\begin{figure}[ht!]
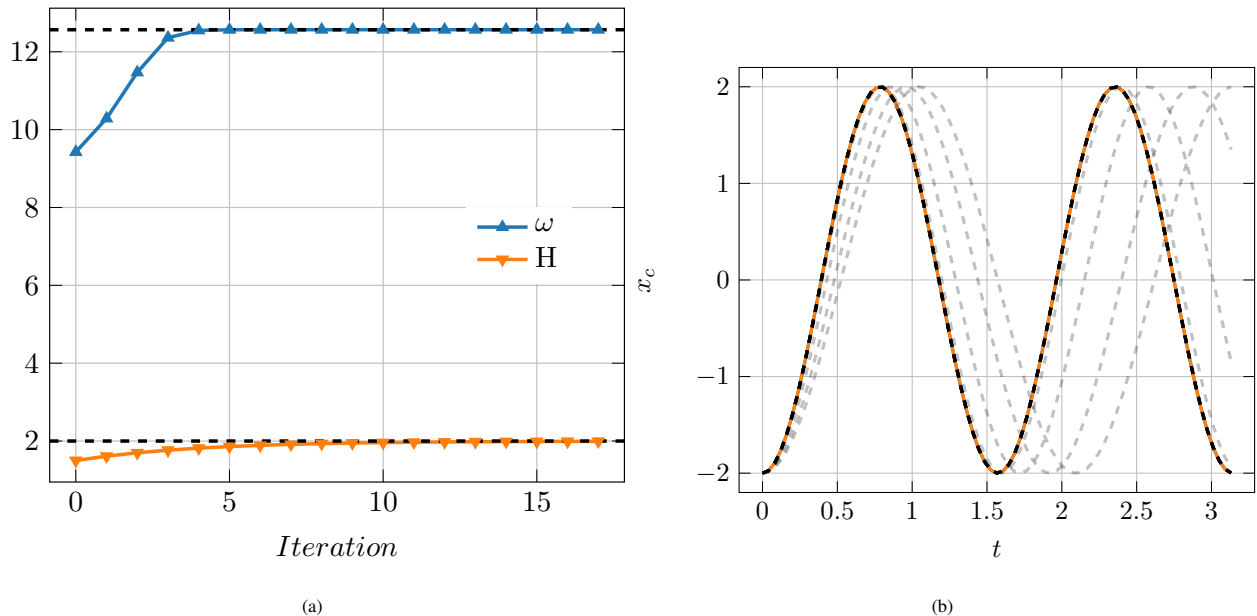

\begin{subfigure}[b]{0.5\linewidth}
\centering
\includefigure{optim_qbx_helix_control}
\caption{}
\end{subfigure}
\hfill
\begin{subfigure}[b]{0.5\linewidth}
\centering
\includefigure{optim_qbx_helix_centroid_x}
\caption{}
\end{subfigure}
\caption{Test 3: (a) Convergence of the control variables $(\omega, H)$ to their
    desired values (dashed). (b) Centroid $\vect{x}_c(t)$ (dashed gray) for the
    droplet starting at $(-R, 0, 0)$ and the desired centroid trajectory
    (dashed orange).}
\label{fig:test3:control}
\end{figure}

For the optimization, we have chosen to start with $(\omega^{(0)}, H^{(0)}) =
(3 \pi, 1.5)$ and $(\omega_d, H_d) = (4 \pi, 2)$. The results of the
optimization can be seen in~\Cref{fig:test3:convergence}. We can see that both
the cost functional and the gradient are significantly decreased during the
optimization process. In particular, the gradient has decreased by about three
orders of magnitude and the cost has decreased by about five orders of magnitude.
The convergence of both quantities is well-behaved throughout the optimization,
when considering the naive line search algorithm used. To better understand
the convergence to the optimal solution, we turn to~\Cref{fig:test3:control}.
We can see here that both the parameters $(\omega, H)$ approach their respective
desired values in only a few iterations. Comparing to~\Cref{fig:test3:convergence},
we can see that once the wavenumber $\omega$ becomes close to $\omega_d$, the
optimization changes its behavior and the error decreases linearly.

\section{Conclusions}
\label{sc:conclusions}

We have developed a methodology for applying optimal control to the Stokes
flow of two-phase immiscible fluids with sharp interfaces. We have considered
systems with constant surface tension, but extensions to other context
can be easily derived (e.g. variable surface tension, gravity, electrostatic).
The focus has been on extending the existing results from~\cite{Fikl2021} to
fully three-dimensional problems containing multiple interacting droplets.

At the theoretical level, the main contribution of this work has been in
deriving a scalar evolution equation for the adjoint transverse field. This is
important for the three-dimensional extension, as it provides a non-trivial
computational cost reduction. Furthermore, the optimal control theory
in the context of moving domains is a very active area of research and the
application to two-phase Stokes flow is an important stepping stone to more
complex scenarios. However, many open questions remain regarding the general
well-posedness of the sensitivity and adjoint equations. Unlike results such
as~\cite{Laurain2021}, the adjoint transverse field~\eqref{eq:adjoint_transverse_field}
does not satisfy a parabolic-type equation, so much less can be said about its
regularity.

We have also presented a numerical scheme based on existing, well-tested
libraries that provides sufficient efficiency and robustness to perform
optimal control of the two-phase flows. Previous work in~\cite{Fikl2021} has
relied on special treatment of each type of singularity and kernel in the
adjoint equations, which was mainly possible due to the one-dimensional interface
(under axisymmetric assumptions). Extending and developing similar methods for
the three-dimensional case is impractical, so we have used the kernel-independent
QBX method as an alternative way of handling the many singularities in the
boundary integral equations.

The applications presented here have been focused on small deformations of
the interface for moderate time horizons. Testing the numerical scheme on
longer time spans and large deformations is the subject of further investigation.
However, this opens up many additional complexities. For example, issues
regarding the general ``reachability'' of some desired state of the system
become important. In this work, we have used controls in $\mathbb{R}^2$ for
the quasi-static case, but for full control of the system a time- and
space-dependent control is likely necessary. Furthermore, we expect the
gradient descent to perform increasingly worse as the interface deforms more
due to the high nonlinearity of the problem. These issues are common to all
shape-based gradient flows and are of great practical and theoretical
interest for future research.

\paragraph{\textbf{Acknowledgements}}
This work was sponsored, in part, by the Office of Naval Research (ONR)
as part of the Multidisciplinary University Research Initiatives (MURI)
Program, under Grant Number N00014-16-1-2617.

\paragraph{\textbf{Declaration of Interests}}
The authors report no conflict of interest.

\appendix
\section{Mesh Adaptation Functional and Gradient}
\label{ax:adaptive}

In this appendix we detail the gradient of the passive mesh stabilization
functional from~\Cref{ssc:methods:timestepping}, as originally given
in~\cite{Zinchenko2013}. We start by stating that motion law is
$\dot{\vect{x}} = \vect{V}$, where the source term is arbitrary. We must then
find a $\vect{V}$ that minimizes
\[
J(\vect{V}) =
\frac{\alpha}{2} \sum_{i} \sum_{j \in \mathcal{N}(i)} \left[
    \od{}{t} \left(
    \frac{r_{ij}^2}{h_{ij}^2} + \frac{h_{ij}^2}{r_{ij}^2}
    \right)
\right]^2
+ \frac{\beta}{2} \sum_{m} \frac{1}{C_m^2}
    \left[\od{C_m}{t}\right]^2,
\]
where we will be assuming that $h_{ij}$ does not depend on the choice
of $\vect{V}$. We start by looking at the first term, for which we have that
\[
r_{ij}^2 = \|\vect{x}_i - \vect{x}_j\|^2,
\]
so
\[
\begin{aligned}
\frac{1}{2} \od{}{t} \left(
\frac{r_{ij}^2}{h_{ij}^2} + \frac{h_{ij}^2}{r_{ij}^2}
\right) = \,\, &
\frac{1}{h_{ij}^2} (\vect{x}_i - \vect{x}_j) \cdot (\vect{V}_i - \vect{V}_j) -
\frac{h_{ij}^2}{r_{ij}^4} (\vect{x}_i - \vect{x}_j) \cdot (\vect{V}_i - \vect{V}_j) \\
=\,\, &
\left(
    \frac{1}{h_{ij}^2} - \frac{h_{ij}^2}{r_{ij}^4}
\right) (\vect{x}_i - \vect{x}_j) \cdot (\vect{V}_i - \vect{V}_j).
\end{aligned}
\]

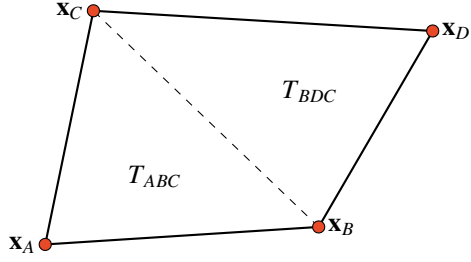
\begin{figure}[ht!]
\centering
\begin{tikzpicture}
\begin{scope}[rotate=-30]
\coordinate (A) at (0, 0);
\coordinate (B) at (3, 2);
\coordinate (C) at (3, 5);
\coordinate (D) at (-1, 3);

\draw[thick] (A) -- (B) -- (C) -- (D) -- cycle;
\draw[dashed] (B) -- (D);

\draw[fill=JCPOrange] (A) circle [radius=0.08] node [left] {$\mathbf{x}_A$};
\draw[fill=JCPOrange] (B) circle [radius=0.08] node [right] {$\mathbf{x}_B$};
\draw[fill=JCPOrange] (D) circle [radius=0.08] node [left] {$\mathbf{x}_C$};
\draw[fill=JCPOrange] (C) circle [radius=0.08] node [right] {$\mathbf{x}_D$};

\node at (0.8, 1.5) {$T_{ABC}$};
\node at (2.0, 3.5) {$T_{BDC}$};
\end{scope}
\end{tikzpicture}

\caption{Numbering of quadrilateral vertices and division intro triangles.}
\label{fig:ax:quadrilateral}
\end{figure}

Then, for the compactness factor $C_m$, we have that
\[
C_m = \frac{|\Sigma_m|}{\sum_{i = 0}^{2^d} |\partial \Sigma_{m, i}|^2},
\]
where $|\Sigma_m|$ denotes the area of the element $\Sigma_m$ and
$|\partial \Sigma_{m, i}|$ are the lengths of its faces. In the case of a
quadrilateral element (for surfaces), the formula reduces to
(see~\Cref{fig:ax:quadrilateral} for notation)
\[
C_m = \frac{|T_{ABC}| + |T_{BDC}|}{r_{AB}^2 + r_{BD}^2 + r_{DC}^2 + r_{CA}^2}.
\]
where $|T_{\cdots}|$ denotes the area of the triangle. We can further express
the area of the triangles through Heron's formula as
\[
\begin{aligned}
|T_{ABC}| =\,\, & \frac{1}{4} \sqrt{
    (r_{AB}^2 + r_{BC}^2 + r_{CA}^2)^2 - 2 (r_{AB}^4 + r_{BC}^4 + r_{CA}^4)
    } = \frac{1}{4} \sqrt{a_{ABC}^2 - 2 b_{ABC}}, \\
|T_{BDC}| =\,\, & \frac{1}{4} \sqrt{
    (r_{BD}^2 + r_{DC}^2 + r_{BC}^2)^2 - 2 (r_{BD}^4 + r_{DC}^4 + r_{BC}^4)
    } = \frac{1}{4} \sqrt{a_{BDC}^2 - 2 b_{BDC}},
\end{aligned}
\]
so we can write
\[
C_m = \frac{\sqrt{a^2_{ABC} - 2 b_{ABC}} + \sqrt{a^2_{BDC} - 2 b_{BDC}}}
            {4 (r_{AB}^2 + r_{BD}^2 + r_{DC}^2 + r_{CA}^2)},
\]
where only squares of the edge lengths appear. We now have the simple derivatives
\[
\pd{a^2_{ABC}}{r_{AB}^2} = 2 (r_{AB}^2 + r_{BC}^2 + r_{CA}^2) = 2 a_{ABC}
\quad \text{and} \quad
\pd{b_{ABC}}{r_{AB}^2} = 2 r_{AB}^2,
\]
with equivalent results for derivatives with respect to $r_{BD}^2, r_{DC}^2$
and $r_{BC}^2$. Note that $r_{BC}^2$ appears in both terms in the numerator,
while it does not appear in the denominator. We can now take the derivative
to the squared variables $r^2_{\cdot \cdot}$ by a simple application of the
chain rule. We take $r_{AB}^2$ and $r_{BC}^2$ as a representative examples,
for which we have
\[
\begin{aligned}
\od{C_m}{r_{AB}^2} = \,\, &
\frac{1}{4 (r_{AB}^2 + r_{BD}^2 + r_{DC}^2 + r_{CA}^2)} \left\{
\frac{r_{BC}^2 + r_{CA}^2 - r_{AB}^2}{|T_{ABC}|}
- 4 C_m
\right\}, \\
\od{C_m}{r_{BC}^2} = \,\, &
\frac{1}{4 (r_{AB}^2 + r_{BD}^2 + r_{DC}^2 + r_{CA}^2)} \left\{
\frac{r_{AB}^2 + r_{CA}^2 - r_{BC}^2}{|T_{ABC}|}
+ \frac{r_{BD}^2 + r_{DC}^2 - r_{BC}^2}{|T_{BDC}|}
\right\}.
\end{aligned}
\]

Finally, the time derivative can be written by the chain rule as
\[
\begin{aligned}
\pd{C_m}{t} = \,\, &
\pd{C_m}{r_{AB}^2} \pd{r_{AB}^2}{t}
+ \pd{C_m}{r_{BD}^2} \pd{r_{BD}^2}{t}
+ \pd{C_m}{r_{DC}^2} \pd{r_{DC}^2}{t}
+ \pd{C_m}{r_{CA}^2} \pd{r_{CA}^2}{t}
+ \pd{C_m}{r_{BC}^2} \pd{r_{BC}^2}{t},
\end{aligned}
\]
where
\[
\frac{1}{2} \pd{r^2_{AB}}{t}
= (\vect{x}_A - \vect{x}_B) \cdot (\vect{V}_A - \vect{V}_B).
\]

We now have all the terms of interest and their derivatives. Determining the
exact formula for the full gradient is straightforward, as all the terms are
simply quadratic in $\vect{V}_i$.

\bibliography{references}

\begin{thebibliography}{10}
\expandafter\ifx\csname url\endcsname\relax
  \def\url#1{\texttt{#1}}\fi
\expandafter\ifx\csname urlprefix\endcsname\relax\def\urlprefix{URL }\fi
\expandafter\ifx\csname href\endcsname\relax
  \def\href#1#2{#2} \def\path#1{#1}\fi

\bibitem{Pozrikidis2001}
C.~Pozrikidis, Interfacial dynamics for {Stokes} flow, Journal of Computational
  Physics 169~(2) (2001) 250--301.
\newblock \href {https://doi.org/10.1006/jcph.2000.6582}
  {\path{doi:10.1006/jcph.2000.6582}}.

\bibitem{Huisken1984}
G.~Huisken, Flow by mean curvature of convex surfaces into spheres, Journal of
  Differential Geometry 20 (1984) 237--266.
\newblock \href {https://doi.org/10.4310/jdg/1214438998}
  {\path{doi:10.4310/jdg/1214438998}}.

\bibitem{Deckelnick2005}
K.~Deckelnick, G.~Dziuk, C.~M. Elliott, Computation of geometric partial
  differential equations and mean curvature flow, Acta Numerica 14 (2005)
  139--232.
\newblock \href {https://doi.org/10.1017/s0962492904000224}
  {\path{doi:10.1017/s0962492904000224}}.

\bibitem{Jameson1988}
A.~Jameson, Aerodynamic design via control theory, Journal of scientific
  computing 3~(3) (1988) 233--260.

\bibitem{Hintermuller2014}
M.~Hintermüller, D.~Wegner, Optimal control of a semidiscrete
  {Cahn--Hilliard--Navier--Stokes} system, {SIAM} Journal on Control and
  Optimization 52 (2014) 747--772.
\newblock \href {https://doi.org/10.1137/120865628}
  {\path{doi:10.1137/120865628}}.

\bibitem{Garcke2019}
H.~Garcke, M.~Hinze, C.~Kahle, Optimal control of time-discrete two-phase flow
  driven by a diffuse-interface model, ESAIM: Control, Optimisation and
  Calculus of Variations 25 (2019) 13.
\newblock \href {https://doi.org/10.1051/cocv/2018006}
  {\path{doi:10.1051/cocv/2018006}}.

\bibitem{Deng2017}
Y.~Deng, Z.~Liu, Y.~Wu, Topology optimization of capillary, two-phase flow
  problems, Communications in Computational Physics 22 (2017) 1413--1438.
\newblock \href {https://doi.org/10.4208/cicp.oa-2017-0003}
  {\path{doi:10.4208/cicp.oa-2017-0003}}.

\bibitem{Prosperetti2009}
A.~Prosperetti, G.~Tryggvason, Computational methods for multiphase flow,
  Cambridge University Press, 2009.

\bibitem{Popinet2009}
S.~Popinet, An accurate adaptive solver for surface-tension-driven interfacial
  flows, Journal of Computational Physics 228~(16) (2009) 5838--5866.

\bibitem{Feppon2019}
F.~Feppon, G.~Allaire, F.~Bordeu, J.~Cortial, C.~Dapogny, Shape optimization of
  a coupled thermal fluid-structure problem in a level set mesh evolution
  framework, SeMA Journal 76~(3) (2019) 413--458.

\bibitem{Bernauer2011}
M.~K. Bernauer, R.~Herzog, Optimal control of the classical two-phase {Stefan}
  problem in level set formulation, {SIAM} Journal on Scientific Computing
  33~(1) (2011) 342--363.

\bibitem{Repke2011}
S.~Repke, N.~Marheineke, R.~Pinnau, Two adjoint-based optimization approaches
  for a free surface {Stokes} flow, {SIAM} Journal on Applied Mathematics
  71~(6) (2011) 2168--2184.

\bibitem{Palacios2012}
F.~Palacios, J.~J. Alonso, A.~Jameson, Shape sensitivity of free-surface
  interfaces using a level set methodology, in: 42nd AIAA Fluid Dynamics
  Conference and Exhibit, 2012.

\bibitem{Laurain2015}
A.~Laurain, S.~W. Walker, Droplet footprint control, {SIAM} Journal on Control
  and Optimization 53 (2015) 771--799.
\newblock \href {https://doi.org/10.1137/140979721}
  {\path{doi:10.1137/140979721}}.

\bibitem{Laurain2021}
A.~Laurain, S.~W. Walker, Optimal control of volume-preserving mean curvature
  flow, Journal of Computational Physics 438 (2021) 110373.
\newblock \href {https://doi.org/10.1016/j.jcp.2021.110373}
  {\path{doi:10.1016/j.jcp.2021.110373}}.

\bibitem{Diehl2020}
E.~Diehl, J.~Haubner, M.~Ulbrich, S.~Ulbrich, Differentiability results and
  sensitivity calculation for optimal control of incompressible two-phase
  {Navier--Stokes} equations with surface tension (2020).
\newblock \href {http://arxiv.org/abs/2003.04971v1}
  {\path{arXiv:2003.04971v1}}.

\bibitem{Kuhl2021}
N.~Kühl, J.~Kröger, M.~Siebenborn, M.~Hinze, Rung, Adjoint complement to the
  {Volume-of-Fluid} method for immiscible flows, Journal of Computational
  Physics 440 (2021) 110411.
\newblock \href {http://arxiv.org/abs/2009.03957v1}
  {\path{arXiv:2009.03957v1}}, \href
  {https://doi.org/10.1016/j.jcp.2021.110411}
  {\path{doi:10.1016/j.jcp.2021.110411}}.

\bibitem{Fikl2021}
A.~Fikl, D.~J. Bodony, Adjoint-based interfacial control of viscous drops,
  Journal of Fluid Mechanics 911 (2021).
\newblock \href {https://doi.org/10.1017/jfm.2020.1013}
  {\path{doi:10.1017/jfm.2020.1013}}.

\bibitem{Klockner2013}
A.~Klockner, A.~Barnett, L.~Greengard, M.~O'Neil, {Quadrature by Expansion}: A
  new method for the evaluation of layer potentials, J. Comput. Phys. 252
  (2013) 332--349.
\newblock \href {https://doi.org/10.1016/j.jcp.2013.06.027}
  {\path{doi:10.1016/j.jcp.2013.06.027}}.

\bibitem{Wala2019}
M.~Wala, A.~Klöckner, A fast algorithm for {Quadrature by Expansion} in three
  dimensions, Journal of Computational Physics 388 (2019) 655--689.
\newblock \href {https://doi.org/10.1016/j.jcp.2019.03.024}
  {\path{doi:10.1016/j.jcp.2019.03.024}}.

\bibitem{Moubachir2006}
M.~Moubachir, J.-P. Zolésio, Moving Shape Analysis and Control: {Applications}
  to Fluid Structure Interactions, Chapman and Hall, 2006.

\bibitem{Luft2020}
D.~Luft, V.~Schulz, Pre-shape calculus: Foundations and application to mesh
  quality optimization (2020).
\newblock \href {http://arxiv.org/abs/2012.09124v1}
  {\path{arXiv:2012.09124v1}}.

\bibitem{Walker2015}
S.~W. Walker, The Shape of Things: {A} Practical Guide to Differential Geometry
  and the Shape Derivative, {SIAM}, 2015.

\bibitem{Allaire2007}
G.~Allaire, M.~Schoenauer, Conception optimale de structures, Springer, 2007.

\bibitem{Pantz2005}
O.~Pantz, Sensibilité de l'équation de la chaleur aux sauts de conductivité,
  Comptes Rendus Mathématique 341~(5) (2005) 333--337.

\bibitem{Gunzburger2003}
M.~D. Gunzburger, Perspectives in Flow Control and Optimization, SIAM, 2003.

\bibitem{Veerapaneni2011}
S.~K. Veerapaneni, A.~Rahimian, G.~Biros, D.~Zorin, A fast algorithm for
  simulating vesicle flows in three dimensions, Journal of Computational
  Physics 230 (2011) 5610--5634.
\newblock \href {https://doi.org/10.1016/j.jcp.2011.03.045}
  {\path{doi:10.1016/j.jcp.2011.03.045}}.

\bibitem{Pozrikidis1992}
C.~Pozrikidis, Boundary Integral and Singularity Methods for Linearized Viscous
  Flow, Cambridge University Press, 1992.

\bibitem{Fikl2020}
A.~Fikl, D.~J. Bodony, Jump relations of certain hypersingular {Stokes} kernels
  on regular surfaces, SIAM Journal on Applied Mathematics 80 (2020)
  2226--2248.
\newblock \href {https://doi.org/10.1137/19m1269804}
  {\path{doi:10.1137/19m1269804}}.

\bibitem{Kress1989}
R.~Kress, Linear Integral Equations, Springer, 1989.

\bibitem{pytential}
A.~Klockner, Collaborators, {pytential}
  \url{https://github.com/inducer/pytential} (2021).

\bibitem{Tornberg2008}
A.~T. Tornberg, L.~Greengard, A fast multipole method for the three-dimensional
  {Stokes} equations, J. Comput. Phys. 227~(3) (2008) 1613--1619.
\newblock \href {https://doi.org/10.1016/j.jcp.2007.06.029}
  {\path{doi:10.1016/j.jcp.2007.06.029}}.

\bibitem{SHTns}
N.~Schaeffer, {SHTns} \url{https://bitbucket.org/nschaeff/shtns} (2021).

\bibitem{Schaeffer2013}
N.~Schaeffer, Efficient spherical harmonic transforms aimed at pseudospectral
  numerical simulations, Geochemistry, Geophysics, Geosystems 14 (2013)
  751--758.
\newblock \href {https://doi.org/10.1002/ggge.20071}
  {\path{doi:10.1002/ggge.20071}}.

\bibitem{Mcclarren2010}
R.~G. McClarren, C.~D. Hauck, Robust and accurate filtered spherical harmonics
  expansions for radiative transfer, Journal of Computational Physics 229
  (2010) 5597--5614.
\newblock \href {https://doi.org/10.1016/j.jcp.2010.03.043}
  {\path{doi:10.1016/j.jcp.2010.03.043}}.

\bibitem{Golub1979}
G.~H. Golub, M.~Heath, G.~Wahba, Generalized cross-validation as a method for
  choosing a good ridge parameter, Technometrics 21 (1979) 215--223.
\newblock \href {https://doi.org/10.1080/00401706.1979.10489751}
  {\path{doi:10.1080/00401706.1979.10489751}}.

\bibitem{Ojala2015}
R.~Ojala, A.~T. Tornberg, An accurate integral equation method for simulating
  multi-phase {Stokes} flows, J. Comput. Phys. 298 (2015).
\newblock \href {https://doi.org/10.1016/j.jcp.2015.06.002}
  {\path{doi:10.1016/j.jcp.2015.06.002}}.

\bibitem{Zinchenko2013}
A.~Z. Zinchenko, R.~H. Davis, Emulsion flow through a packed bed with multiple
  drop breakup, Journal of Fluid Mechanics 725 (2013) 611--663.
\newblock \href {https://doi.org/10.1017/jfm.2013.197}
  {\path{doi:10.1017/jfm.2013.197}}.

\bibitem{Kropinski2001}
M.~C.~A. Kropinski, An efficient numerical method for studying interfacial
  motion in two-dimensional creeping flows, Journal of Computational Physics
  171 (2001) 479--508.
\newblock \href {https://doi.org/10.1006/jcph.2001.6787}
  {\path{doi:10.1006/jcph.2001.6787}}.

\bibitem{Zabarankin2013}
M.~Zabarankin, I.~Smagin, O.~M. Lavrenteva, A.~Nir, Viscous drop in
  compressional {Stokes} flow, Journal of Fluid Mechanics 720 (2013) 169--191.
\newblock \href {https://doi.org/10.1017/jfm.2013.6}
  {\path{doi:10.1017/jfm.2013.6}}.

\bibitem{Burdakov2019}
O.~Burdakov, Y.~Dai, N.~Huang, Stabilized {Barzilai-Borwein} method, Journal of
  Computational Mathematics 37 (2019) 916--936.
\newblock \href {https://doi.org/10.4208/jcm.1911-m2019-0171}
  {\path{doi:10.4208/jcm.1911-m2019-0171}}.

\bibitem{Stone1989}
H.~A. Stone, L.~G. Leal, Relaxation and breakup of an initially extended drop
  in an otherwise quiescent fluid, Journal of Fluid Mechanics 198 (1989)
  399--427.

\end{thebibliography}

\end{document}